\documentclass[showpacs, showkeys,twocolumn,preprintnumbers,floatfix,amsmath,amssymb]{revtex4}

\usepackage{amsmath,amsthm}
\usepackage{graphicx}% Include figure files
\usepackage{dcolumn}% Align table columns on decimal point
\usepackage{bm}% bold math
\usepackage{color}
\usepackage{epsfig}
\usepackage{multirow}
\usepackage{epstopdf}

\begin{document}

\preprint{Preprint}

\title{Self-energy-modified Poisson-Nernst-Planck equations: WKB approximation \\ and  finite-difference approaches}

\author{Zhenli Xu} \email{xuzl@sjtu.edu.cn}
\author{Manman Ma} \email{mmm@sjtu.edu.cn}
\author{Pei Liu} \email{hgliupei1990@sjtu.edu.cn}
\affiliation{Department of Mathematics, Institute of Natural Sciences, and MoE Key Lab of Scientific and Engineering Computing, Shanghai Jiao Tong University, Shanghai 200240, China}


\date{\today}

%%%%% Begin Abstract %%%%%%%%%%%
\begin{abstract}
We propose a modified Poisson-Nernst-Planck (PNP) model to investigate charge transport in electrolytes of inhomogeneous dielectric environment. The model includes the ionic polarization due to the dielectric inhomogeneity and the ion-ion correlation. This is achieved by the self energy of test ions through solving a generalized Debye-H\"uckel (DH) equation. We develop numerical methods for the system composed of the PNP and DH equations. Particularly, towards the numerical challenge of solving the high-dimensional DH equation, we developed an analytical WKB approximation and a numerical approach based on the selective inversion of sparse matrices. The model and numerical methods are validated by simulating the charge diffusion in electrolytes between two electrodes, for which effects of dielectrics and correlation are investigated by comparing the results with the prediction by the classical PNP theory. We find that, at the length scale of the interface separation comparable to the Bjerrum length, the results of the modified equations are significantly different from the classical PNP predictions mostly due to the dielectric effect. It is also shown that when the ion self energy is in weak or mediate strength, the WKB approximation presents a high accuracy, compared to precise finite-difference results.

\end{abstract}
%%%%% end %%%%%%%%%%%

%%%%% AMS/PACs/Keywords %%%%%%%%%%%

%\ams{}

%%%% maketitle %%%%%
\maketitle

 %%%% Start %%%%%%

\section{Introduction}

The charge transport in fluids under confinements or around objects of nanometer length scale has been of growing interest in a lot of physical and biological systems \cite{Schoch:RMP:08,Daiguji:CSR:10}; for example, in the study of colloidal separation and self-assembly, nanoparticles at liquid-liquid interfaces, electrochemical energy devices and membrane ionic channels. When the length scale of confinements is comparable with the characteristic lengths of electrolytes (e.g., the Bjerrum length of the solvent $\ell_B$, and the Debye length of the electrolyte $\ell_D$), complex electrostatic phenomena such as the charge inversion and the like-charge attraction have been often observed experimentally. These phenomena are beyond mean-field theoretical explanations and motivate many challenges to computational and modelling communities \cite{FPP+:RMP:2010}.

In the vicinity of a charged surface, counterions are attracted, forming a screening region called electric double layer. The structure of the double layer plays a key role in nanoscale interface sciences \cite{Levin:RPP:2002,WK+:N:2011,FPP+:RMP:2010,Messina:JPCM:2009,wan2014:PRX}.  The ions in the external diffuse layer is Coulombic, where the electric potential decays exponentially with the characteristic Debye length. The ions in the internal Stern layer are condensed, where the electric potential does not obey the exponential rule due to the tangential ion correlation and other short-range interactions. If the medium contacting the electrolyte is a dielectric of low permittivity, the dielectric mismatch leads to an induced-charge potential, repelling the counterions and thus creating a depletion zone near the surface. This effect could play an important role in many phenomena of nanoscale systems, and is also attracting much attention in theoretical and simulation study (see  \cite{Messina:JCP:2002,HL:SM:2008,JKN+:PRL:2008,WM:JPCB:2010,Bakhshandeh:PRL:2011,DL:JPCM:2012a,GXX:JCP:2012,Xu:PRE:2013,WangRui:JCP:13,ZO:PNAS:13} to mention a few recent literature).

In the mean-field description of ion structure and transport in the double layer, the Poisson-Boltzmann (PB) and Poisson-Nernst-Planck (PNP) theories have been considered. The PNP is beyond the PB because the PNP describes the charge dynamics and thus has been widely used for studying ion transport in nanopores and nanochannels; for example, open ion channels in cell membranes \cite{CE:BJ:93, E:JMB:96, E:JMB:99, HCE:JSC:01}. The system of equations has also been widely applied in continuum theories of narrow channels by coupling with the Stokes equations to model nanofluidics \cite{daiguji:NL:2005,Daiguji:CSR:10}. The PNP equations are also known as ``drift-diffusion equations" and have much use in semiconductors \cite{MRS:S:90}. Asymptotic analysis on the PNP equations has been made to understand properties of charge diffusion of electrochemical energy systems \cite{BTA:PRE:04, SB:JFM:04, SN:SIAM:09}. From the aspect of numerical analysis, various numerical methods have been proposed \cite{KCG:BJ:99, CCK:BJ:00, ZCW:JCP:11, WZCX:SIAM:12, Flavell:JCE:14,dyrka2013Proteins} to solve the nonlinear equations more accurately and efficiently and to capture specific dynamical behaviors of the non-equilibrium system.

The classical PNP theory neglects the ion-ion correlation, excluded volume and image charge effects in the double layer, significantly changing the local structure and the far-field properties of electrolytes. When large surface charges or applied potentials
are involved, the nonmonotonic differential capacitance of electrodes can not be predicted without these effects \cite{LSR:JPC:08}. Analysis of current-concentration curves  \cite{MRA:BJ:03} also showed significantly different trends between the experimental data \cite{BTH:BJ:98} and that predicted by the PNP theory. Moreover, it has been found that the mean-field theory fails in explaining, even qualitatively,
phenomena of the long range attractions between two surfaces in electrolyte \cite{KS:PRL:95}.
The comparison between the PNP theory and particle-based Brownian dynamics simulations \cite{CKC:BJ:00} for cylindrical pores of varying sizes demonstrated the invalidity of the mean-field theories when the cylindrical radius is less than $2\ell_D$, evidencing that the aforementioned effects should be accounted for in the continuum modeling.

Different versions of modified PB/PNP theory have been proposed by including such as steric effects \cite{storey2008pre,KBA:PRE:07, HTLE:JPC:12,LZ:BJ:2011,EHL:JCP:2010,liu2012jdde} or the dielectric self energy \cite{nadler2003pre, CKC:BJ:03, GKCN:JPC:04, HEL:CMS:11}.
A useful approach of remedying the mean-field theory is to replace the mean potential by a better approximation of the potential of mean force, i.e., to include the ignored effects by correcting the mean potential by the self energy of mobile ions.
The remedy to modify the mean potential may improve the mean-field theory a lot by including ionic correlations and image charges to capture many-body physics; see, e.g., \cite{Luo06,WangRui:JCP:13}.
The shortcoming is that the self energy is very difficult to obtain, either by molecular dynamics simulations or by solving high-dimensional equations from field-theory calculations. In this work, we follow the idea of Gaussian variational field theory \cite{podgornik1989jcp,NO:EPJE:2000,NO:EPJE:2003} which models the self energy as the self-Green's function. A self-energy modified PNP model is derived (Sec. II) and the Green's function is described by the generalized Debye-H\"uckel (DH) equation nonlinearly depending on the mean ion distributions. The resulted system of partial differential equations is then composed of three equations: the Nernst-Planck equations for the dynamics of charges, the Poisson equation for the mean potential, and the generalized DH equation for the Green's function. The self energy is solved efficiently by some analytical and numerical approximations, and then the modified PNP equations are solved by finite difference algorithm.

In Section III, we discuss the dimensionless formulation of the modified PNP equations, by introducing two ratios of length scales, $\epsilon=\ell_D/L$ and $q=\ell_B/L$, where $L$ is the characteristic length of confinements. Then we developed the analytical WKB  (Wentzel-Kramers-Brillouin) approximation and the numerical finite-difference approximation for the DH equation. The WKB approach attempts to express the self energy as an explicit formula by asymptotic approximation assuming the parameter $\epsilon$ is small. The finite-difference numerical method is more expensive, but can be tackled by using selective inversion for sparse symmetric and positive definite matrix. In Section 4, our numerical results show the WKB approximation is in a good agreement with the difference method when $L$ is bigger than a few Bjerrum lengths. This is a positive evidence that the use of the WKB approximation gives a satisfactory accuracy in the self-energy calculation, while avoiding expensive high-dimensional numerical calculations.

Our numerical examples place much focus on the dielectric effect of electrodes. This effect is believed to be important in many systems as aforementioned, and may play important role in many-body interaction between colloids \cite{WangRui:JCP:13,ZO:PNAS:13}. Certainly, electrostatic interaction at solid-liquid interfaces is much more complex than the simplified picture. One important effect is the variable dielectric permittivity of solvent, which has been found to capture many key phenomena observed experimentally \cite{BGN:PRL:2011,BN:JPC:13}. The modified PNP model does include the treatment of this effect with varying dielectric constant and excluded volumes of test ions by the DH equation; Eq. \eqref{dh} below. This treatment will highly increase the computational cost compared with the present model, thus, we remain our model here for a dielectrically homogeneous electrolyte bounded by sharp interfaces.

\section{Governing equations}

The transport of charged particles in an electrolyte is often described by the Nernst-Planck equation (also called the Smoluchowski equation) which states that the time derivative of the ion density function is composed of a diffusion contribution and an advection contribution due to the potential energy $U_i$ of the ions,
\begin{equation}
\frac{\partial c_i}{\partial t} = \nabla\cdot D_i\left(\nabla c_i+\beta c_i\nabla U_i \right), ~~~i=1,\cdots, N,
\end{equation}
where $c_i$ is the concentration of the ions of species $i$, $D_i$ is its diffusion constant,  $\beta=1/k_BT$ where $k_B$ is the Boltzmann constant and $T$ is the temperature.

In the mean-field description of electrostatics, the potential energy $U_i$ takes the mean potential energy $U_i=z_ie\Phi$,
and the electric potential $\Phi$ is governed by the Poisson equation,
\begin{equation}
-\varepsilon_0\nabla\cdot \varepsilon \nabla\Phi=\rho_fe+\sum_i z_ie c_i,
\end{equation}
where $z_i$ is the valence, $\rho_f e$ is the fixed charge, $\varepsilon_0$ is the vacuum dielectric permittivity and $\varepsilon(\textbf{r})$ is the relative dielectric permittivity of the medium. The coupling system between the Nernst-Planck equation and the Poisson equation is called the Poisson-Nernst-Planck (PNP) equations.

The mean-field nature of the PNP ignores the induced-charge effect (or image-charge effect) of dielectric discontinuity and many-body ion correlation, which plays an important role in a lot of electrostatic phenomena. Even at the weak-coupling limit, the induced-charge effect is significant and should be taken into account. In order to include these effects, the potential energy of transported particles $U_i$ can be expressed as the mean potential energy plus a correction,
\begin{equation}
U_i=z_ie\Phi+\frac{1}{2}z_i^2e^2 u_i,
\end{equation}
where $u_i$ is the self energy of a unit test ion of the $i$th transported species. Given the mean potential $\Phi$, an accurate description of $u_i$ is to include all many-body interaction with the test ion, i.e., the potential of mean force, which for example can be done by molecular dynamics simulations \cite{Luo06}, or by a reaction-field formulation \cite{CKC:BJ:03}. In the self-consistent Gaussian field approximation \cite{NO:EPJE:2000,NO:EPJE:2003,Wang:PRE:2010}, this quantity can be defined through a Green's function $G(\textbf{r},\textbf{r}')$, described by the generalized Debye-H\"uckel (DH) equation,
\begin{equation}
-\varepsilon_0 \nabla\cdot  \varepsilon_i'(\textbf{r},\textbf{r}') \nabla G +2I(\textbf{r},\textbf{r}') G = \delta(\textbf{r}-\textbf{r}'), \label{dh}
\end{equation}
and $u_i$ is then the self Green's function limit,
\begin{equation}
u_i=\lim_{\mathbf{r'} \rightarrow\mathbf{r}} [G(\textbf{r},\textbf{r}') - G_0(\textbf{r},\textbf{r}')],
\end{equation}
where $G_0=1/(4\pi\varepsilon_0\varepsilon_\mathrm{eff} |\textbf{r}-\textbf{r}'|)$ is the Green's function in free space to remove the invariable singularity, and $I$ is the local ionic strength,
which describes the screening effect by the surrounding ions of the test ion, and where the ionic concentrations are determined by the Nernst-Planck equation. The prime in the permittivity function $\varepsilon_i'$ describes the function has been locally modified due to the excluded volume of the test ion,
\begin{equation}
\varepsilon_i'(\mathbf{r},\mathbf{r}')=\left\{ \begin{array}{ll}
\varepsilon_\mathrm{eff},~~|\mathbf{r}-\mathbf{r}'|<a_i,\\
\varepsilon(\mathbf{r}),~~\hbox{otherwise},
\end{array}\right.
\end{equation}
where $a_i$ is the ionic radius, and $\varepsilon_\mathrm{eff}$ is the effective dielectric permittivity inside the ionic cavity. We see the ionic size effect is taken into account by assuming that the ionic cavity is inaccessible to mobile ions. Consequently, in Eq. \eqref{dh}, $I$ is expressed as,
\begin{equation}
I(\textbf{r},\textbf{r}')=\left\{ \begin{array}{ll}  ~~0,~~~~ |\mathbf{r}-\mathbf{r}'|<a_i,\\
\frac{1}{2}\beta e^2 \sum_i z_i^2 c_i,~~\hbox{otherwise}.
\end{array}\right.
\end{equation}
This modification is particularly useful to remove the self-energy singularity when the permittivity of solvent is space-dependent \cite{cherepanov2003bj,BN:JPC:13}, but greatly increases numerical difficulty. To avoid this difficulty, we will not discuss this size effect in this work, and simply take the limit $a_i\rightarrow 0$, and leave the algorithm development for a future study.

In equilibrium, the zero ion flux of each species leads to the following equality,
\begin{equation*}
D_i\left(\nabla c_i+\beta c_i\nabla U_i \right)=0.
\end{equation*}
Solving this equation gives an explicit formula for the equilibrium ion density,
\begin{equation*}
c_i=c_{i0} e^{-\beta U_i},
\end{equation*}
where $c_{i0}$ is constant determined by the chemical potential. Substituting the expression into the right side of the Poisson equation, we obtain a modified Poisson-Boltzmann equation,
\begin{equation}
-\varepsilon_0\nabla\cdot\varepsilon \nabla\Phi=\rho_f e+\sum_i z_ie c_{i0} e^{-\beta U_i}.
\end{equation}
Together with the generalized DH equation \eqref{dh}, it has been studied by Avdeev and Martynov \cite{AM:CJU:1986} using the Debye closure of the BBGKY chain, Netz and Orland \cite{NO:EPJE:2000,NO:EPJE:2003} using Gaussian variational field theory, and in much recent work (see \cite{BAA:JCP:2012,Yaroshchuk:AIS:2000,HL:SM:2008,LLP+:PRE:2002} to mention a few).

\section{Charge dynamics in the presence of planar surfaces}

\begin{figure}[htbp!]
    \centering
    \includegraphics[width=.9\linewidth]{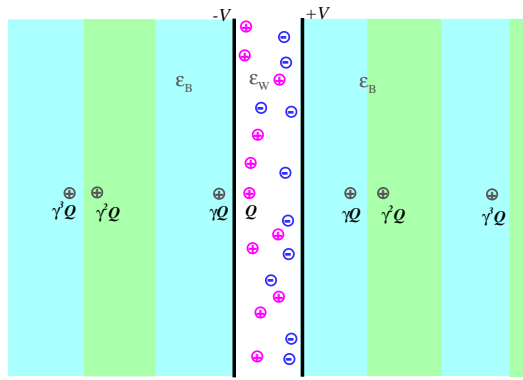}
    \caption{(Color online) Schematic illustration of the electrolyte between two dielectric interfaces with applied voltages. When $\varepsilon_B<\varepsilon_W$, the coefficient $\gamma>0$ and the mobile ions near the interfaces are repelled by image charges.}
    \label{fig:schem}
\end{figure}

In order to understand the self-energy effects for the charge transport, we consider an electrolyte with 1:1 salt  between two parallel planar electrodes at $x=\pm L$ (see Fig. \ref{fig:schem}), which is a simple model of electrochemical systems. We study the case of sharp dielectric permittivity $\varepsilon$ which takes the water permittivity $\varepsilon_W$ for $|x|<(1+\xi)L$ with a small $\xi>0$, and the alternative value $\varepsilon_B$ for the electrodes outside. The use of a small separation between the dielectric interface and the electrode avoids the self-energy divergence near the boundary. The sharp interfaces remain, providing the induced-charge effect to the mobile ions in the electrolyte.

\subsection{Dimensionless equations, boundary and initial conditions}

Let $\ell_B= \beta e^2/(4\pi \varepsilon_0\varepsilon_W)$
be the Bjerrum length at the water solvent, at which distance the interaction energy of two unit point charges is $k_BT$.
We take $L$ as the reference length scale,  $D_0$ as a typical diffusion constant, and $c_0$ as a typical ion condensation.
We assume ions have uniform diffusion constant and define the another length scale $\ell_D=1/\sqrt{8\pi\ell_B c_0}$. We note, when $c_0$ is the bulk ion concentration for symmetric monovalent salt, $\ell_D$ is the Debye screening length.

The following dimensionless parameters and variables are defined: $\widetilde{\textbf{r}}=\textbf{r}/L$,  $\widetilde{t}=t D_0/L\ell_D,$ $\widetilde{c}_i= c_i/c_0$ and $\widetilde{\rho}_f=\rho_f/c_0$, $\widetilde{D}_i=D_i/D_0=1$, $\widetilde{\varepsilon}=\varepsilon/\varepsilon_W$, $\widetilde{\Phi}=\beta e \Phi$, $\widetilde{G}= \beta e^2 G$, $\widetilde{G}_0=\beta e^2 G_0$.  We drop the tildes of all new variables.  Then the modified PNP and generalized DH equations are,
\begin{eqnarray}
&&\frac{\partial c_i}{\partial t} = \epsilon \frac{\partial}{\partial x} \left[\frac{\partial}{\partial x} c_i+ c_i\frac{\partial}{\partial x} \left(z_i\Phi+\frac{1}{2} z_i^2 u\right) \right], i=1,2,~~~~~~ \\
&&-2\epsilon^2\frac{\partial^2}{\partial x^2} \Phi=\rho_f+ \sum_i z_i c_i,  \\
&&-\epsilon^2\nabla \cdot \varepsilon \nabla G + \frac{1}{2}\sum_i z_i^2 c_i G = 4\pi q \epsilon^2\delta(\textbf{r}-\textbf{r}'),   \\
&&G_0=\frac{q}{|\textbf{r}-\textbf{r}'|}, \\
&&u=\lim_{\mathbf{r'} \rightarrow\mathbf{r}} [G(\textbf{r},\textbf{r}') - G_0(\textbf{r},\textbf{r}')], ~~-1<x<1,
\end{eqnarray}
where $\epsilon=\ell_D/L$ and $q=\ell_B/L$ are two length-scale parameters. We see $\epsilon$ and $2\epsilon^2$ describe the effective diffusion constant of the Nernst-Planck equation, and dielectric permittivity of the Poisson and generalized DH equations, and $q$ represents the charge of the test ion, i.e., the strength of the self energy.

The Nernst-Planck and the Poisson equations are defined on a finite interval $-1<x<1$, while the generalized DH equation is defined on the whole three-dimensional domain with implied interface conditions at $x=\pm(1+\xi)$. We assume completely blocking electrodes \cite{BTA:PRE:04} imposed with a time-varying external potential, $V_\pm$ in dimensionless unit, on the electrodes. Then, at $x=\pm 1$, the ionic fluxes should vanish and the electric potential is a fixed time function. This boundary condition leads to a global zero flux at the steady state, i.e., the equilibrium solution. The current-voltage relations for channel problems correspond to different boundary conditions and will be explored in a future publication under the self-energy modified PNP model. Concerning the generalized DH equation, the Green's function should be solved in an infinite domain, i.e., the decaying boundary condition. Therefore, we have the following boundary conditions for concentrations, potential and the Green's function,
\begin{eqnarray}
&&\frac{\partial}{\partial x} c_i+ c_i\frac{\partial}{\partial x} \left(z_i\Phi+\frac{1}{2} z_i^2 u\right)=0, ~~x=\pm 1, \\
&&\Phi  \pm \eta \epsilon \frac{\partial \Phi}{\partial x}= V_\pm, ~~x=\pm 1, \\
&&G\rightarrow 0, ~~\hbox{when}~|\textbf{r}| \rightarrow \infty.
\end{eqnarray}
In this work, we take $\eta=0$ to use the Dirichlet condition for the potential. The Robin boundary condition, i.e., $\eta\neq0$, is often used to account for the dielectric electrodes in literature \cite{BTA:PRE:04,Flavell:JCE:14} by considering the tight counterion adsorption at the Stern layer.
In the Green's function equation, the dielectric permittivity is discontinuous at interfaces at $x=\pm(1+\xi)$. We find as a result of using a positive $\xi$, the higher-order singularity of the self energy can be removed when $x'\rightarrow \pm 1$. The dimensionless dielectric permittivity is then $\varepsilon=1$ for $|x|\leq (1+\xi)$, and $\varepsilon_B/\varepsilon_W$ elsewhere.

Initially, it is assumed that the ions are in equilibrium without applying the external potential which starts to function at $t=0$. Therefore, we can set constant initial ionic concentrations, $c_i(x,0)=c_i^0$, under electric-neutral constraint $\sum_i z_i c_i^0=0$. For monovalent salt we are studying, $c_i^0=1$ for $i=1$ and 2. In addition, we assume there is no fixed charge $\rho_f=0$ in the system, though fixed charges play important roles in many biological and nanofluidic devices.

Now we have completed initial and boundary conditions for the PNP model. The system properties are determined by the ionic species, the applied external potential, and the ratios of length scales $\epsilon$ and $q$.

\subsection{Discretization of the PNP equations}

The Nernst-Planck and the Poisson equations are approximated by finite-difference discretizations. Let $\Delta t$ and $\Delta x$ be the time and space steps, and denote the $n$th time and $k$th space grids by $t^n=n\Delta t$ and $x_k=-1+k\Delta x$, respectively.
The second-order semi-implicit time-stepping scheme is employed for the Nernst-Planck equation,
\begin{equation}
\frac{c_i^{n+1}-c_i^n}{\Delta t} = \epsilon \frac{\partial}{\partial x} \left[\frac{\partial}{\partial x} c_i^{n+\frac{1}{2}}+ c_i^{n+\frac{1}{2}}\frac{\partial}{\partial x} \left(\frac{3}{2}U_i^n-\frac{1}{2}U_i^{n-1}\right) \right],
\end{equation}
where $U_i=z_i\Phi+\frac{1}{2} z_i^2 u,$ and $c_i^{n+\frac{1}{2}}=\frac{c_i^n+c_i^{n+1}}{2}.$ Here the scheme of diffusion term is a central difference, and that of the advection term is an extrapolation for approximating $U_i^{n+\frac{1}{2}}$. By this discretization, we gain the second-order accuracy in time and the benefit of avoiding nonlinear iterations thanks to the use of the explicit approximation for the nonlinear advection coefficient. The stability condition mainly depends on the implicitly-discretized diffusion term, and thus the grid sizes can be $\Delta t \varpropto\Delta x$. Since it is a three-point scheme in time, we should use the backward scheme for $U_i$ at the initial step $n=0$.

Concerning the spatial discretization, because the dielectric function $\varepsilon=1$ is constant in interval $-1<x<1$ and the ionic concentrations are smooth, we could rewrite the advection term as $\partial_x (c_i \partial_x U_i)=(\partial_x c_i)(\partial_x U_i)+ c_i \partial_{xx} U_i$. Standard three-point central differences are used for the first and the second space derivatives of $c_i$ and $U_i$ for the internal points. In discretizing the boundary conditions, two ghost points outside the boundaries are introduced to obtain second-order central approximations. For the Poisson equation with given $c_i^{n+1}$, the mean potential at $(n+1)$th time step can be simply obtained by central discretization with the boundary condition.

\subsection{Solution of the generalized DH equation}

Different from the Nernst-Planck and the Poisson equations, the discretization for the generalized DH equation is not trivial. We rewrite the generalized DH equation as a space-dependent-coefficient form, which is required to solve in each time step,
\begin{equation}
-\nabla\cdot \varepsilon \nabla G + \kappa(x)^2 G = 4\pi \delta(\textbf{r}-\textbf{r}'), ~~~-\infty<x<\infty,
\end{equation}
where the generalized inverse Debye length, $\kappa(x)=\sqrt{\sum_i z_i^2 c_i^{n+1}/(2\epsilon^2)}$ between two interfaces and zero otherwise. Here we set $q=1$ without loss of generality. Solving this equation gives the Green's function at time $t^{n+1}$. This equation is more difficult due to higher dimensions of the Green's function.

We will introduce two approaches. The WKB approximation is often adopted to find approximate Green's function in the presence of electrolytes and interfaces due to its analytical nature \cite{BAA:JCP:2012,WangRui:JCP:13}, which avoids numerical solution of the high-dimensional problem. We propose a simple WKB expression by improving the idea of Buff and Stillinger \cite{Buff:JCP:63} which has been used recently for studying the double layer interaction by Wang and Wang \cite{WangRui:JCP:13}. This approach results in an explicit formula of the self energy with clear physical significance. In the second approach we propose a numerical approximation with finite difference, which could reach any desired accuracy by varying the grid sizes, and so we limit the use of WKB to a pure analytical formulation.

\subsubsection{WKB approximation}

In the WKB approximation introduced by Buff and Stillinger \cite{Buff:JCP:63} for one-interface problem, the Green's function is first exactly found in the case of $\kappa(x)$ being zero, then the approximate Green's function takes the screened Coulomb potential for each image charges using the local ion concentration for the inverse screening length. In the presence of two interfaces, the Green's function of the salt-free solution is a series of image charges by the reflection between two interfaces, as illustrated in Fig. \ref{fig:schem}. For finite $\kappa$,  by the WKB, we have the Green's function as,
\begin{equation}
G(\mathbf{r},\mathbf{r}')=\sum_{\ell=-\infty}^\infty \frac{\gamma^{|\ell|}e^{-\kappa'(x,x')r_\ell}}{r_\ell},
\end{equation}
where $r_\ell$ is the distance between $\mathbf{r}$ and the $\ell$th image charge located at $\left(\ell D+(-1)^\ell x',y',z'\right),$ $D=2(1+\xi)$ is the separation of interfaces and $\gamma$ describes the jump in dielectrics  $\gamma=\frac{\varepsilon_W-\varepsilon_B}{\varepsilon_W+\varepsilon_B}$. The function of the inverse screening length is the average between $x$ and $x'$, i.e.,
$$\kappa'(x,x')=\frac{1}{x-x'} \int_{x'}^x \kappa(s) ds.$$
By subtracting the free-space Green's function and taking the self Green function limit, $\tilde{\kappa}(x,x')=\kappa(x)$ and the self energy $u$ is then found, for $-1\leq x\leq 1,$
\begin{eqnarray}
&&u(x)=-\kappa(x)+\sum_{\ell=2,4,\cdots} \frac{2\gamma^{\ell}e^{-\kappa(x)\ell D}}{\ell D} \nonumber \\
&&~ +\sum_{\ell=1,3,\cdots}\gamma^\ell\left[
\frac{e^{-\kappa(x)(\ell D+2x)}}{\ell D+2x}+\frac{e^{-\kappa(x)(\ell D- 2x)}}{\ell D-2x}
\right],~~~~ \label{wkb}\end{eqnarray}
where the first term $u_\mathrm{loc}=-\kappa(x)$ is the contribution from the local ions, and the two series are the contribution from image charges. It can be observed that, if $\varepsilon_B<\varepsilon_W,$ then $\gamma>0$ and we notice that the image charges are repulsive to the ions. If $\varepsilon_B>\varepsilon_W,$ then the image charges with odd $\ell$ are attractive, but those of even $\ell$ are repulsive.

The WKB formulation \eqref{wkb} is derived from the perturbative expansion with small $\kappa$, and is also accurate for large $\kappa$ because the strong screening leads to weak image potentials. Thus the formulation is aysmptotically exact for both small and large $\kappa$ limits. For the intermediate $\kappa$, the WKB approximation can be considered as an interpolation of the two limits, which, however, may produce poor prediction for the self energy of ions near the interfaces due to inaccurate estimation of the local contribution $-u_\mathrm{loc}$. In the following, we propose an improved approximation.

We remain the form of the image charge series, but modifying the local contribution $\widetilde{\kappa}(x)=-\widetilde{u}_\mathrm{loc}$ by the similar technique of Born series (truncated at the first order) widely known in the field of quantum scattering.  Consider the following two Green's functions $G_1(\mathbf{r},\mathbf{r}')$ and $G_0(\mathbf{r},\mathbf{r}')$ which satisfy,
\begin{eqnarray}
&-\nabla^2 G_1+\kappa(x)^2 G_1=4\pi\delta(\mathbf{r}-\mathbf{r}'), &\label{g1} \\
&-\nabla^2 G_0=4\pi\delta(\mathbf{r}-\mathbf{r}'). & \label{g2}
\end{eqnarray}
Let $\delta G=G_1-G_0$, then $\tilde{u}_\mathrm{loc}$ can be approximately,
\begin{equation}
\widetilde{u}_\mathrm{loc} = \lim_{\mathbf{r'} \rightarrow\mathbf{r}} \delta G(\mathbf{r},\mathbf{r}').
\end{equation}
Subtracting Eq. \eqref{g1} by Eq. \eqref{g2}, we have the following approximate equation for $\delta G$,
\begin{equation}
-\nabla^2 \delta G = -\kappa(x)^2 \widetilde{G}_1(\mathbf{r},\mathbf{r}'), \label{diff}
\end{equation}
where in the right side the assumption $\widetilde{G}_1=e^{-\kappa(x) |\mathbf{r}-\mathbf{r}'|}/|\mathbf{r}-\mathbf{r}'|$ is applied by considering $\widetilde{G}_1$ is a perturbation of $G_1$.

Since Eq. \eqref{diff} is homogeneous in $y$ and $z$. Now we take polar-coordinate Fourier transform in these two coordinates to get,
\begin{equation}
-\partial_{xx} \widehat{\delta G} + \omega^2 \widehat{\delta G}=\frac{-1}{b(x)} \kappa(x)^2 e^{-b(x)|x-x'|},
\end{equation}
where $b(x)=\sqrt{\omega^2+\kappa(x)^2}$.
The solution of this equation for $x\in[-D/2,D/2]$ can be expressed in an integral form by using one-dimensional Green's function,
\begin{eqnarray}
&&\widehat{\delta G}= \int_{-D/2}^{D/2}  \frac{1}{2\omega} e^{-\omega|x-x''|} \cdot
\frac{-1}{b(x'')} \kappa(x'')^2 e^{-b(x'')|x''-x'|}dx'' \nonumber \\
&&~~~~=\int_{-D/2}^{D/2}  \frac{  -\kappa(x'')^2 }{2b(x'')\omega} e^{-(\omega|x-x''|+b(x'')|x''-x'|)} dx''.
\end{eqnarray}
Since $\epsilon\ll1$, $\kappa\gg 1$ and $b$ is a big quantity, we have the asymptotic,
\begin{eqnarray}
&&\widehat{\delta G}(\omega; x'\rightarrow x)\approx\frac{ -\kappa(x)^2 }{2b(x)\omega}\int_{-D/2}^{D/2}  e^{- (\omega+b(x))|x-x''|} dx'' \nonumber\\
&&~~ =-\frac{ \kappa(x)^2 [2-e^{-(\omega+b)(D/2-x)}-e^{-(\omega+b)(D/2+x)}]}{2b\omega(\omega+b)}.~~~~~~
\end{eqnarray}
Let $\rho=\sqrt{(y-y')^2+(z-z')^2}$, then we obtain explicit expression of the inverse Fourier transform, by setting $y'\rightarrow y$ and $z'\rightarrow z$, which is,
\begin{equation}
\widetilde{u}_\mathrm{loc} \approx -\kappa(x)  c(x),
\end{equation}
where
%\begin{widetext}\begin{eqnarray}
%&c(x)& =\lim_{\rho\rightarrow0}\int_0^\infty \frac{[2-e^{-(\omega+b)(D/2-x)}-e^{-(\omega+b)(D/2+x)}]}{2b\omega(\omega+b)} J_0(\rho \omega)\omega d\omega \nonumber \\
%&&\approx 1+\kappa\int_0^\infty \frac{[-e^{-(\omega+\kappa)(D/2-x)}-e^{-(\omega+\kappa)(D/2+x)}]}{2(\omega+\kappa)^2}d\omega
%    \nonumber \\
%&&=1+\frac{1}{2}\left \{-e^{-\kappa(D/2-x)} -e^{-\kappa(D/2+x)} +\kappa(D/2-x) \Gamma [0,\kappa(D/2-x) ]+\kappa(D/2+x)\Gamma [0,\kappa(D/2+x) ]\frac{}{}\right \}.
%\end{eqnarray}\end{widetext}
\begin{eqnarray}
&c(x)&=\lim_{\rho\rightarrow0}\int_0^\infty \frac{[2-e^{-(\omega+b)(D/2-x)}-e^{-(\omega+b)(D/2+x)}]}{2b\omega(\omega+b)}  \nonumber \\
&&~~~~~~~~\displaystyle\cdot J_0(\rho \omega)\omega d\omega \frac{}{} \nonumber \\
&&\approx 1+\kappa \int_0^\infty \frac{[-e^{-(\omega+\kappa)(D/2-x)}-e^{-(\omega+\kappa)(D/2+x)}]}{2(\omega+\kappa)^2}d\omega
    \nonumber \\
&&=1+\frac{\mathfrak{F}(\kappa(x)(D/2+x))+ \mathfrak{F}(\kappa(x)(D/2-x))}{2},
\end{eqnarray}
and
\begin{equation}
\mathfrak{F}(\eta)= \eta \Gamma[0, \eta]-e^{- \eta}.
\end{equation}
Here, $J_0$ is the Bessel function, and $\Gamma[\alpha,z]\doteq\int_z^\infty t^{\alpha-1} e^{-t}dt$ is the incomplete Gamma function.

Then we could replace $\kappa(x)$ in Eq. \eqref{wkb} by $\widetilde{\kappa}(x)=-\widetilde{u}_\mathrm{loc}$ to obtain an improved version of the WKB approximation,
\begin{eqnarray}
&&u(x)=-\widetilde{\kappa}(x)+\sum_{\ell=2,4,\cdots} \frac{2\gamma^{\ell}e^{-\widetilde{\kappa}(x)\ell D}}{\ell D} \nonumber \\
&&~+\sum_{\ell=1,3,\cdots}\gamma^\ell\left[
\frac{e^{-\widetilde{\kappa}(x)(\ell D+2x)}}{\ell D+2x}+\frac{e^{-\widetilde{\kappa}(x)(\ell D- 2x)}}{\ell D-2x}\right], ~~~~ \label{wkb2}\end{eqnarray}
and now the image-charge effect is included. The improvement from $\kappa$ to $\widetilde{\kappa}$ lies in taking into account the edge effect near the boundaries, i.e., the solvent is in confinement between two interfaces. With it, the ion interaction near interfaces is strengthened due to the weaker screening.

\subsubsection{Finite difference approximation}

The numerical solution of the Green's function is usually difficult due to high dimensionality -- it is a function of both the source and field coordinates. We will use the properties of the geometric symmetry of the considered system and only the self Green's function being required. These two properties allow us to develop very efficient solver to get the self energy, which has been coupled with the modified Poisson-Boltzmann equation to simulate equilibrium charged systems \cite{XuMaggs:arXiv:13}. We briefly overview the algorithm below.

In order to reduce the dimensions, the polar symmetric Fourier transform with respect to $y-y'$ and $z-z'$ is first applied. Let $\widehat{G}$ and $\widehat{G}_0$ be the frequency correspondences of $G$ and $G_0$, and let $\omega$ be the frequency. The Fourier transform of the generalized DH equation is a two-dimensional equation of $x$ and $x'$ for each $\omega$,
\begin{equation}
\left[-\frac{\partial}{\partial x} \varepsilon \frac{\partial}{\partial x}  + \varepsilon \omega^2 +\kappa(x)^2\right] \widehat{G}(\omega; x,x') = 2 \delta(x-x'). \label{freq}
\end{equation}
Similarly, we perform Fourier transform for the free-space Green's function equation, $-\nabla^2 G_0=4\pi q \delta(\textbf{r}-\textbf{r}')$, for which we get,
\begin{equation}
\left(-\frac{\partial^2}{\partial x^2} + \omega^2 \right) \widehat{G}_0(\omega; x,x') = 2 \delta(x-x').
\end{equation}
To numerically solve the equations, we discretize the derivatives by central differences for $\widehat{G}$ and $\widehat{G}_0$.
The Dirac delta function is approximated by Kronecker delta,
$\delta(x_j-x_k) \approx \delta_{jk}/\Delta x.$
It should be noticed that the free-space Green's function should be numerically approximated in order to cancel the numerical singularity of the Green's function.
Then for each $\omega$, the frequency Green's functions are obtained by finding the inverse of the coefficient matrices of linear algebra system by discretizing Eq. \eqref{freq}. Then the image and correlation self energy is given by the inverse Fourier transform,
\begin{equation}
u_j=\int_0^\infty \left[ \widehat{G}_{jj}(\omega) - \widehat{G}_{0,jj}(\omega) \right] \omega d\omega, \label{integ}
\end{equation}
where $\widehat{G}_{jj}(\omega)$ and $\widehat{G}_{0,jj}(\omega)$ are diagonal elements of the approximate frequency Green's functions. This equation is approximated by numerical integration.

We remark that the direct inversion of a matrix is expensive, and we should apply some sparse inversion technique to find the diagonals of an inverse matrix; for literature, see  \cite{LYM+:ATMS:2011,LYL+:SJoSC:2011,PM:PRA:2009,George:SJNA:1973,Davis:SIAM:06}.

\section{Numerical results}

\begin{figure*}[t!]
    \centering
    \includegraphics[width=.32\linewidth]{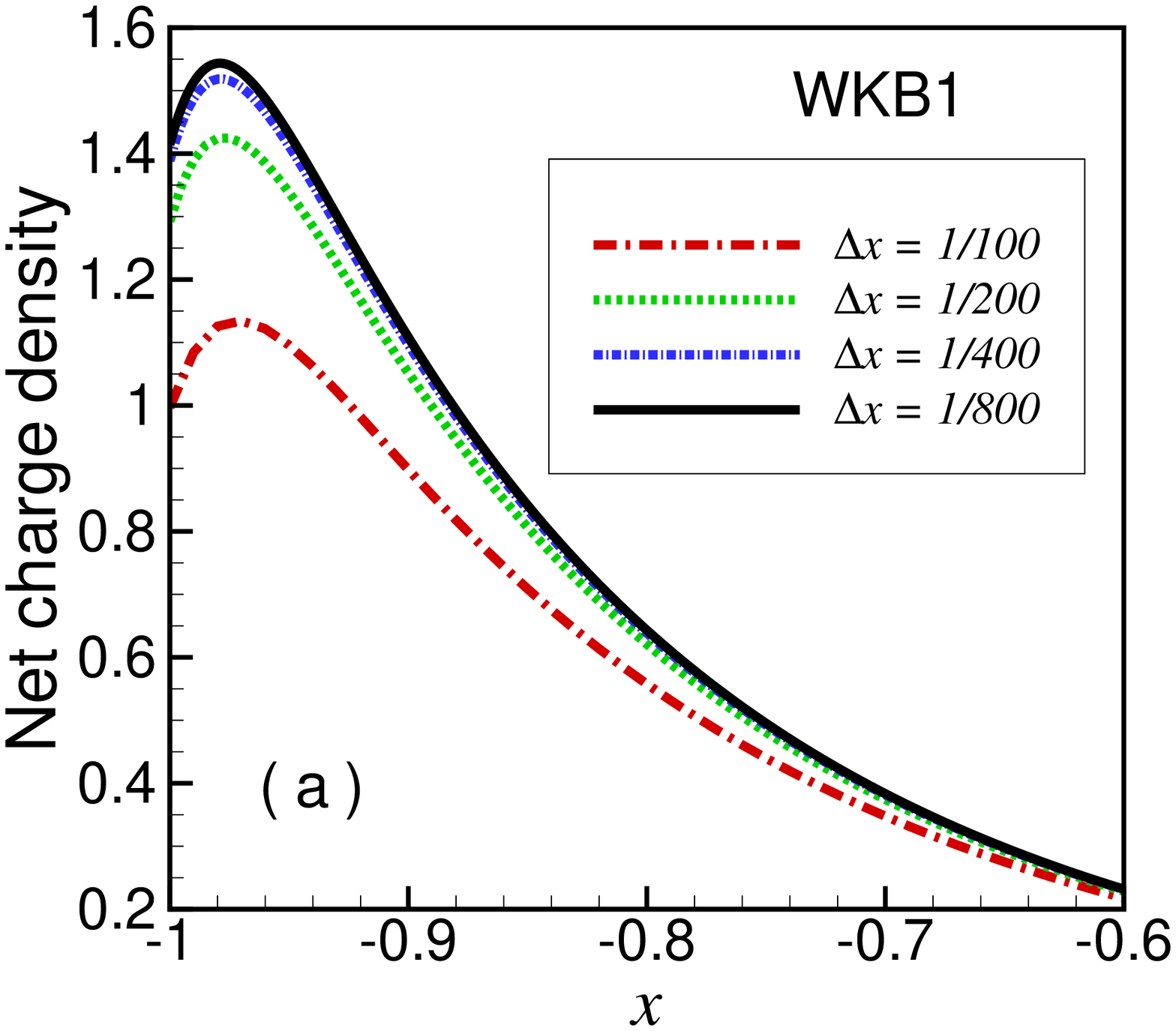}
\includegraphics[width=.32\linewidth]{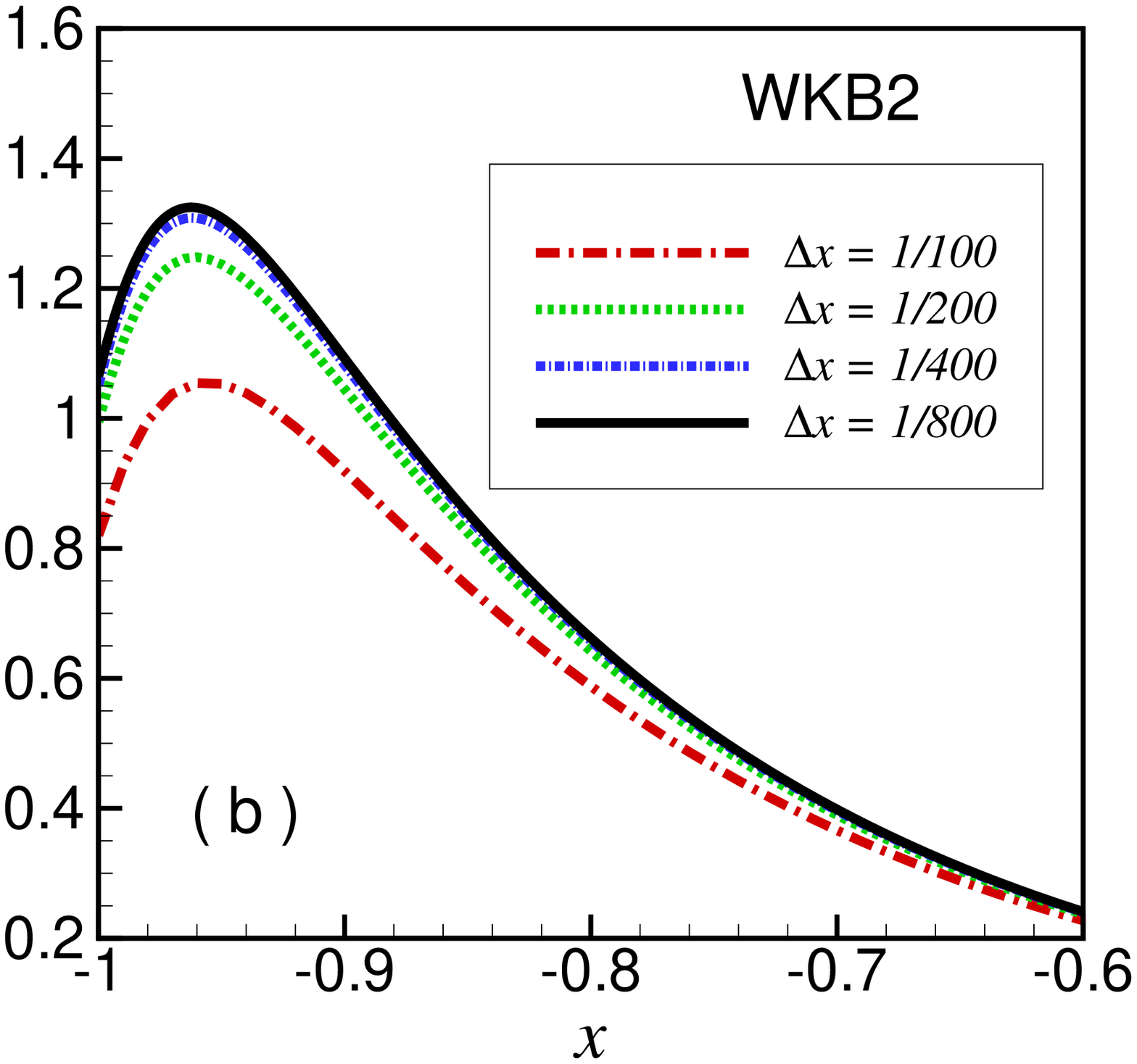}
\includegraphics[width=.32\linewidth]{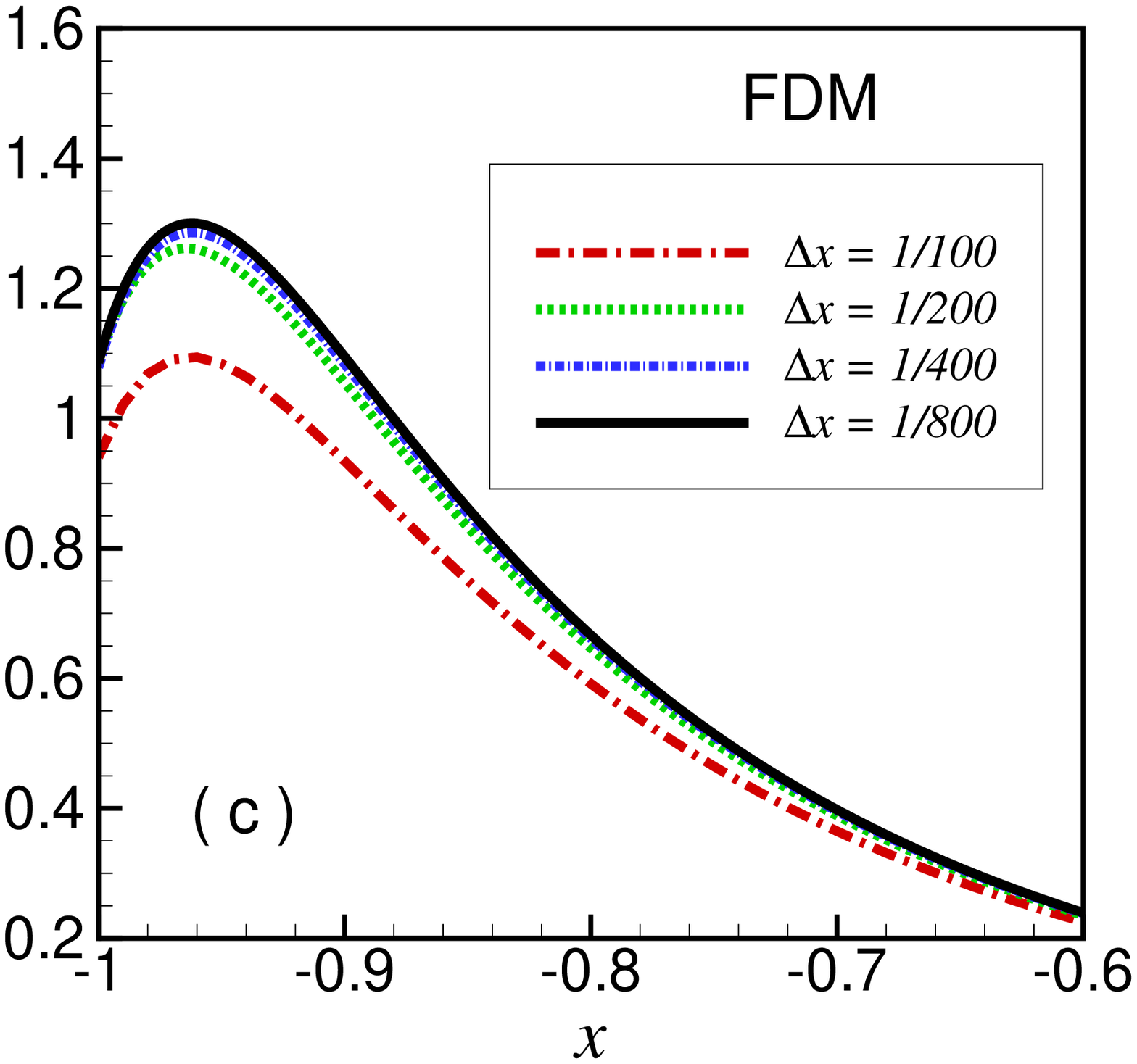}
    \caption{(Color online) Convergence of numerical methods with WKB1, WKB2, and FDM approximations to the Green's function. Net charge densities in interval $x\in[-1,-0.6]$ with $\Delta x=1/100, 1/200, 1/400$ and $1/800$ are plotted.}
    \label{fig:accuracy}
\end{figure*}

In this section, we present numerical results by performing calculations of monovalent electrolytes. Two species of ions with valence $z_{1,2}=\pm 1$ are included, and initially the dimensionless ion densities set  $c_{1,2}=1$. The separation between the dielectric interface and the boundary takes $\xi=0.06$, and thus $D=2.12$. In all calculations, the parameters $\epsilon=0.2$ and electrode voltage $V_\pm=\pm1$ (correspondingly, $\pm 25.6 mV$) are fixed, and different strengths of $q$ and $\varepsilon_B/\varepsilon_W$ are investigated. Since the Debye length $\ell_D$ (typically, $\sim 1-100 nm$) is the characteristic thickness of the electric double layer, this choice of $\epsilon$ does not introduce much interaction between two double layers and continuum theory is thus considered adequate to describe the charge dynamics \cite{Daiguji:CSR:10}, and we will see that introducing self-energy contribution significantly modifies the system quantities at the length scale.

The WKB and finite-difference numerical approaches for the generalized DH equation are compared. Since we have two WKB approaches, we label the results of different approaches by ``WKB1", ``WKB2" and ``FDM", where WKB1 is based on expression \eqref{wkb} and WKB2 is the improved version Eq. \eqref{wkb2}. Also we present the results of classical PNP, denoted by ``PNP" (corresponding to $q=0$), to show the difference between the PNP and the modified model. The image charge series in WKB formulations are truncated at $\ell=10$, large enough to ensure that the truncation error is negligible.
In FDM approximation, the infinite integral Eq. \eqref{integ} is approximated through a cutoff at a frequency $\omega=\Omega$ to become an integral over $[0, \Omega]$ and a variable transformation $\omega=e^{\mu v}-1$ for $\mu>0$. We take $\mu=1$, $\Omega=1024$ and 16 quadrature points in the calculations. These parameters provide high accuracy for the integration.

\subsection{Convergence of numerical methods}

In the first example, numerical schemes for the modified PNP equations with both the WKB and FDM approximations are tested. We set the parameter $q=0.2$ and the dielectric ratio  $\varepsilon_B/\varepsilon_W=0.05$, and compute the results up to time $T=2$. For a water solvent at room temperature, $\varepsilon_W\approx 80$ and $\ell_B\approx 0.7 nm$, the value of $q=0.2$ means the distance between electrodes is $L \sim 3.5 nm$ and is also physically interesting for studies of nanoscale devices. This choice of big $q$ is convenient to observe the convergence of the schemes. The dielectric constant $\varepsilon_B=4$ is typical for membranes or other materials. The time step takes $\Delta t=\Delta x/2$ with varying space grid sizes $\Delta x=1/100, 1/200, 1/400$ and $1/800$.

Fig. \ref{fig:accuracy} presents the results of the spatial distribution of the net charge density, $z_1c_1(x,T)+z_2c_2(x,T)$, with the three approaches and the four grid sizes. It is shown that the error of the maximum values between $\Delta x\leq 1/400$ and $1/800$ is less then $2\%$. By taking the results of the finest grid ($\Delta x=1/800$) as the reference, it is observed that the error is also decreased in a factor of $\sim 4$ if $\Delta x$ is halved, demonstrating all approaches are self-convergent with the second-order accuracy.

The three approaches all predict a non-monotonic curves near the interfaces. Near the cathode, the net charge is first raising to a maximum, then monotonically decays to zero at $x=0$. This is due to $\varepsilon_B<\varepsilon_W$ thus the self-image charges repel the mobile ions, and in agreement with Monte Carlo simulations \cite{Messina:JCP:2002,GX:PRE:2011,WM:JCP:2009}. In contrast, The PNP ignores the polarization effect, and thus always predicts monotonic net charge density (shown in next section). By comparing the three approaches, the charge density predicted from the WKB1 is much higher than those from the FDM and the WKB2. The maximums of the solid black lines of Fig. \ref{fig:accuracy} (a)-(c) are 1.54, 1.32, and 1.30, respectively. As the FDM with the fine mesh is considered very accurate, it is concluded that the WKB1 overestimates the charge density near the interface, and lowers down the induced-charge effect. The WKB2 curves are in good agreements with the FDM results, showing a high accuracy of the analytical approximation.

\subsection{Effect of self-energy strength}

To investigate the effect of different $q$,  two groups of parameters are adopted: four self-energy strengths $q=0,~ 0.05, ~0.1$ and 0.2, and four time snapshots $T=0.2,~ 0.5,~ 2$ and 10. When $q=0$, the generalized DH equation is switched off and the model reduces to the classical PNP, and thus the PNP results are also compared to investigate how the self energy influences the results. The space grid size takes $1/800$. Other parameters remain the same as the previous example.

\begin{figure*}[htbp!]
    \centering
    \includegraphics[width=.4\linewidth]{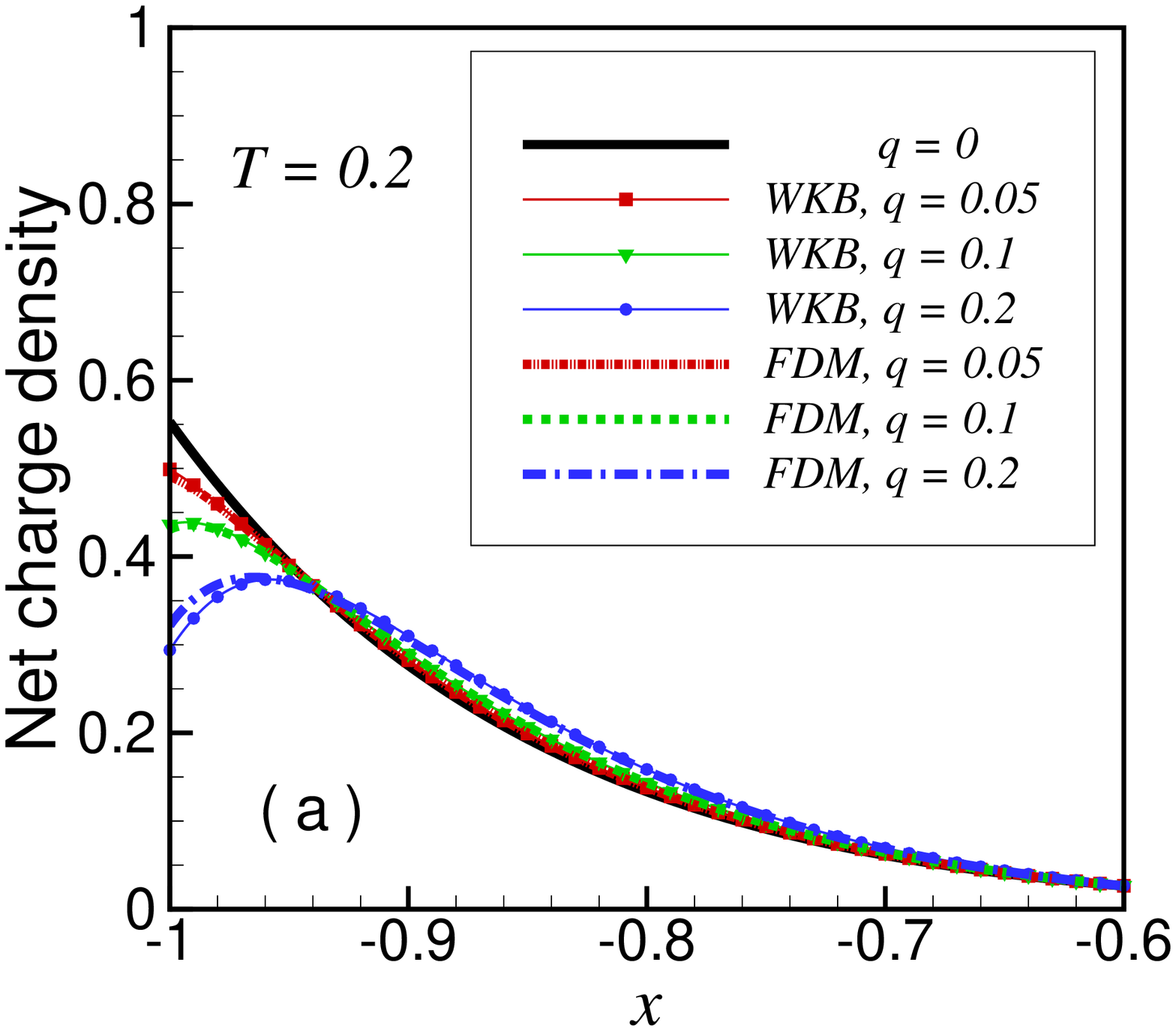}\includegraphics[width=.4\linewidth]{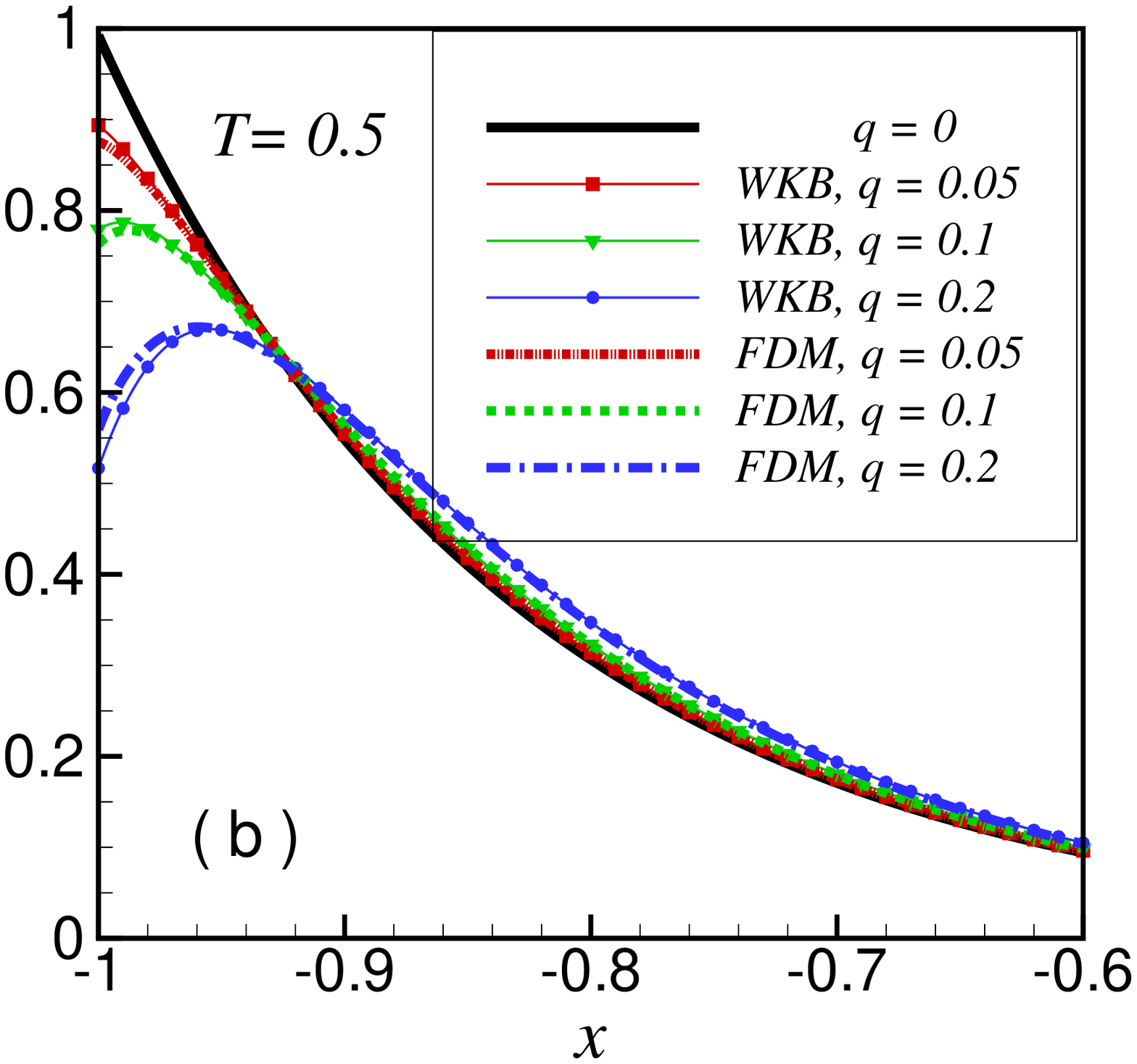} \\
    \includegraphics[width=.4\linewidth]{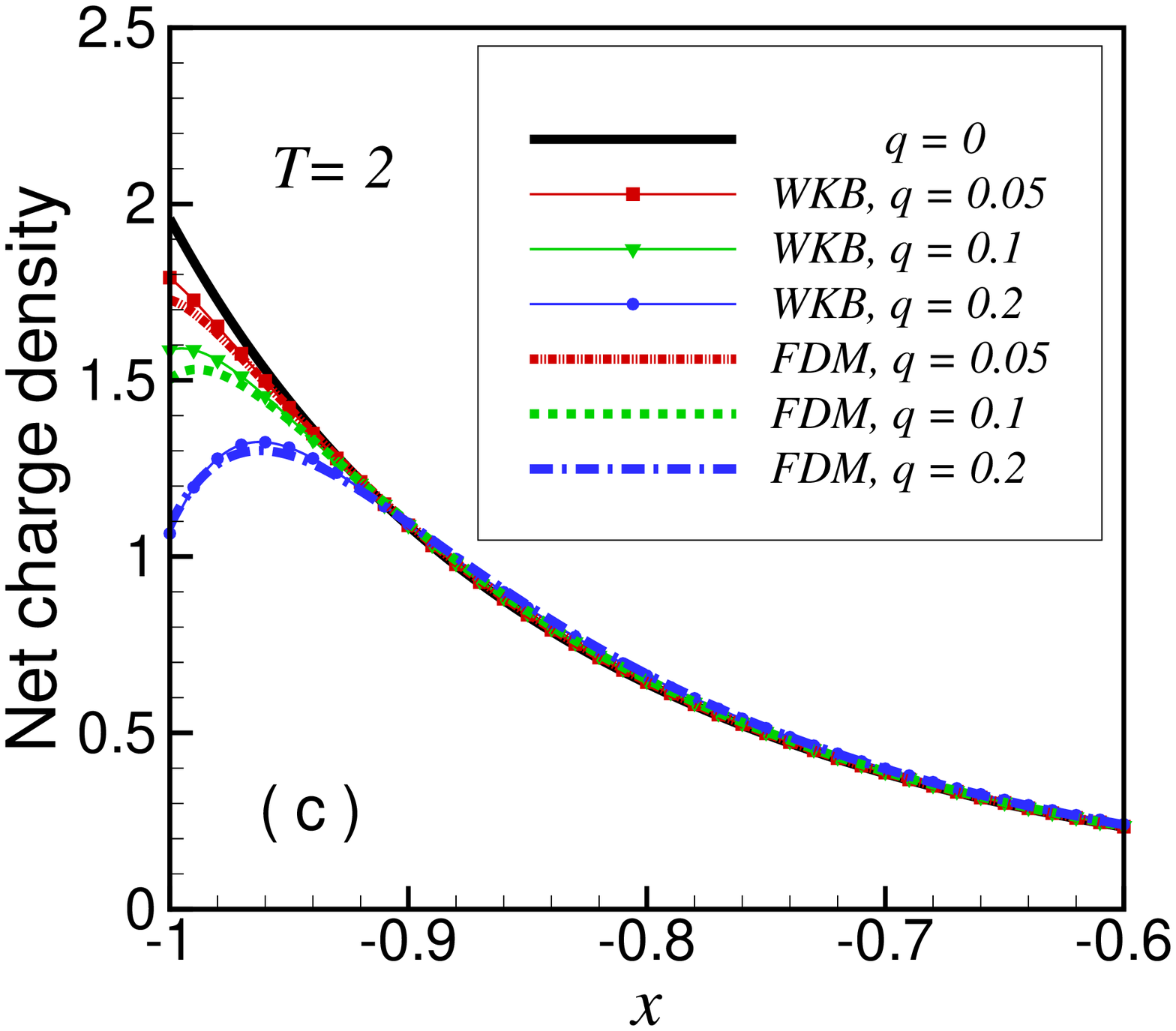}\includegraphics[width=.4\linewidth]{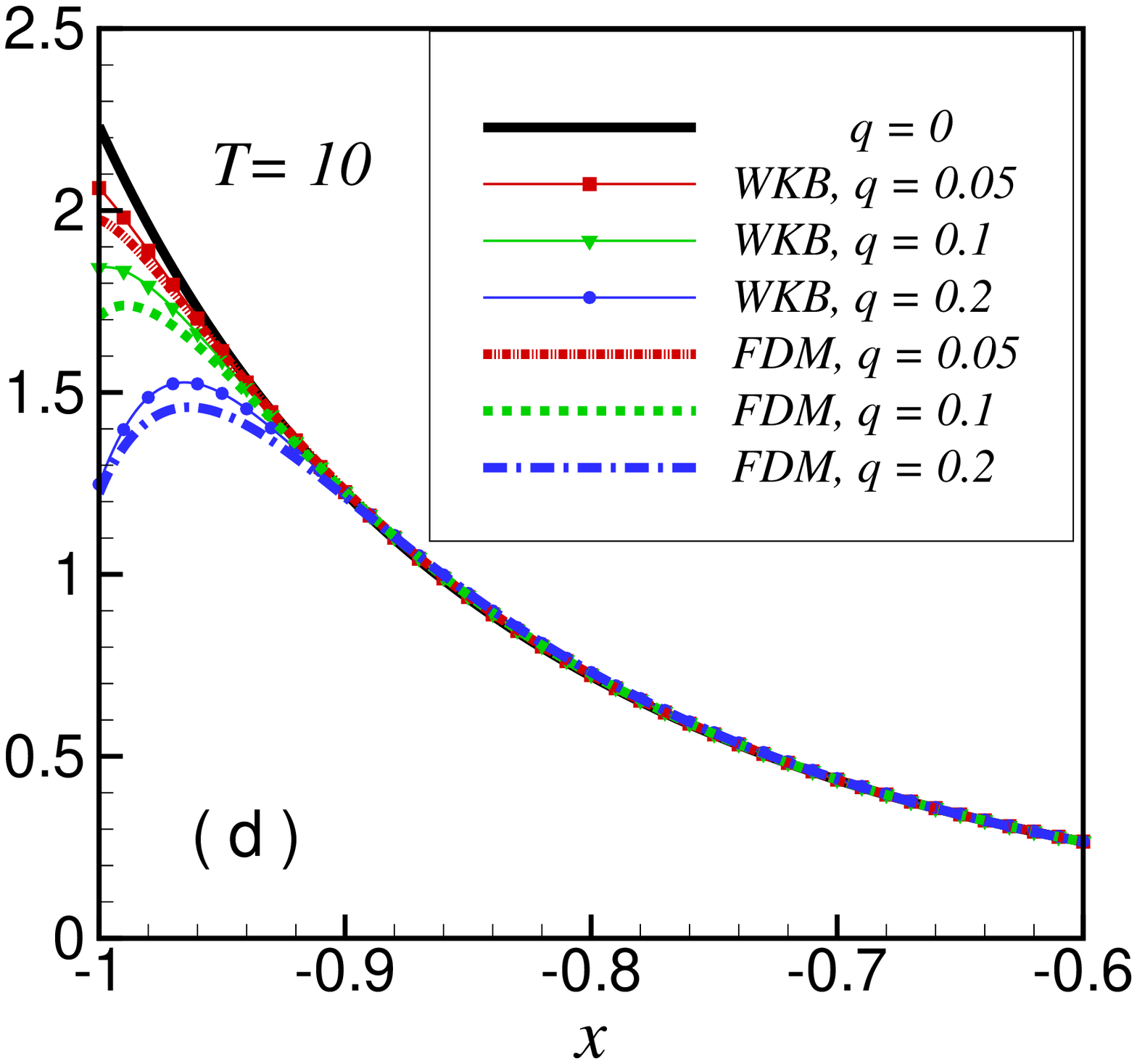}
    \caption{(Color online) Comparison of the net charge density for different $q$ values between WKB2 and FDM  at four different time snapshot $T=0.2, 0.5, 2$ and 10. $q=0$ is the classical PNP.}
    \label{fig:qT}
\end{figure*}

Fig. \ref{fig:qT} plots the net-charge-density results by the PNP, WKB2 and FDM for the four time snapshots. Again, both WKB2 and FDM are in good agreement. It is observed that image repulsion is strengthened with the increase of $q$. For $q=0.2$, there is an obvious maximum at $x\approx-0.95$. Far-field curves are overlapping at equilibrium state ($T=10$).

\begin{figure*}[htbp!]
   \includegraphics[width=.4\linewidth]{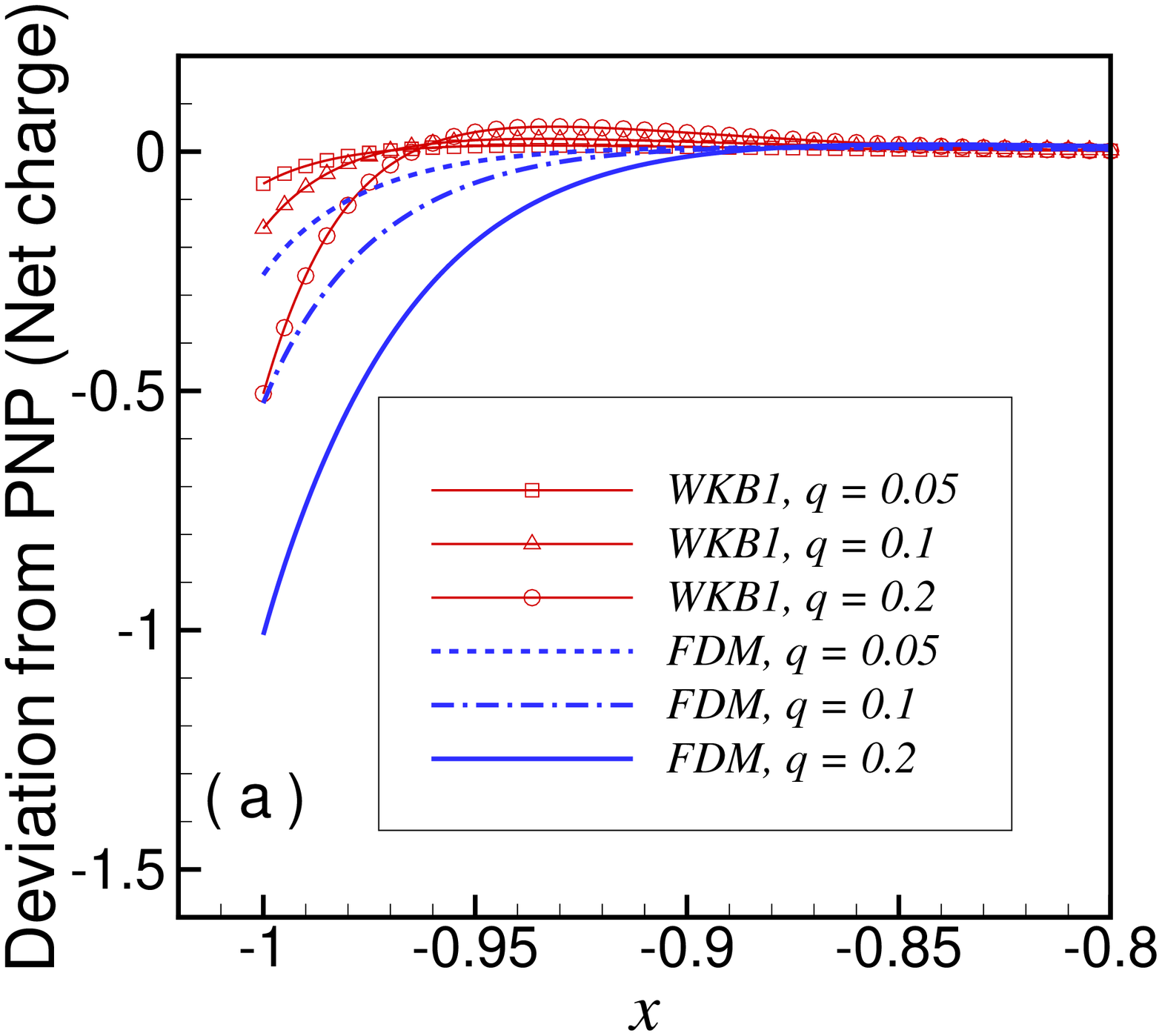}  \includegraphics[width=.4\linewidth]{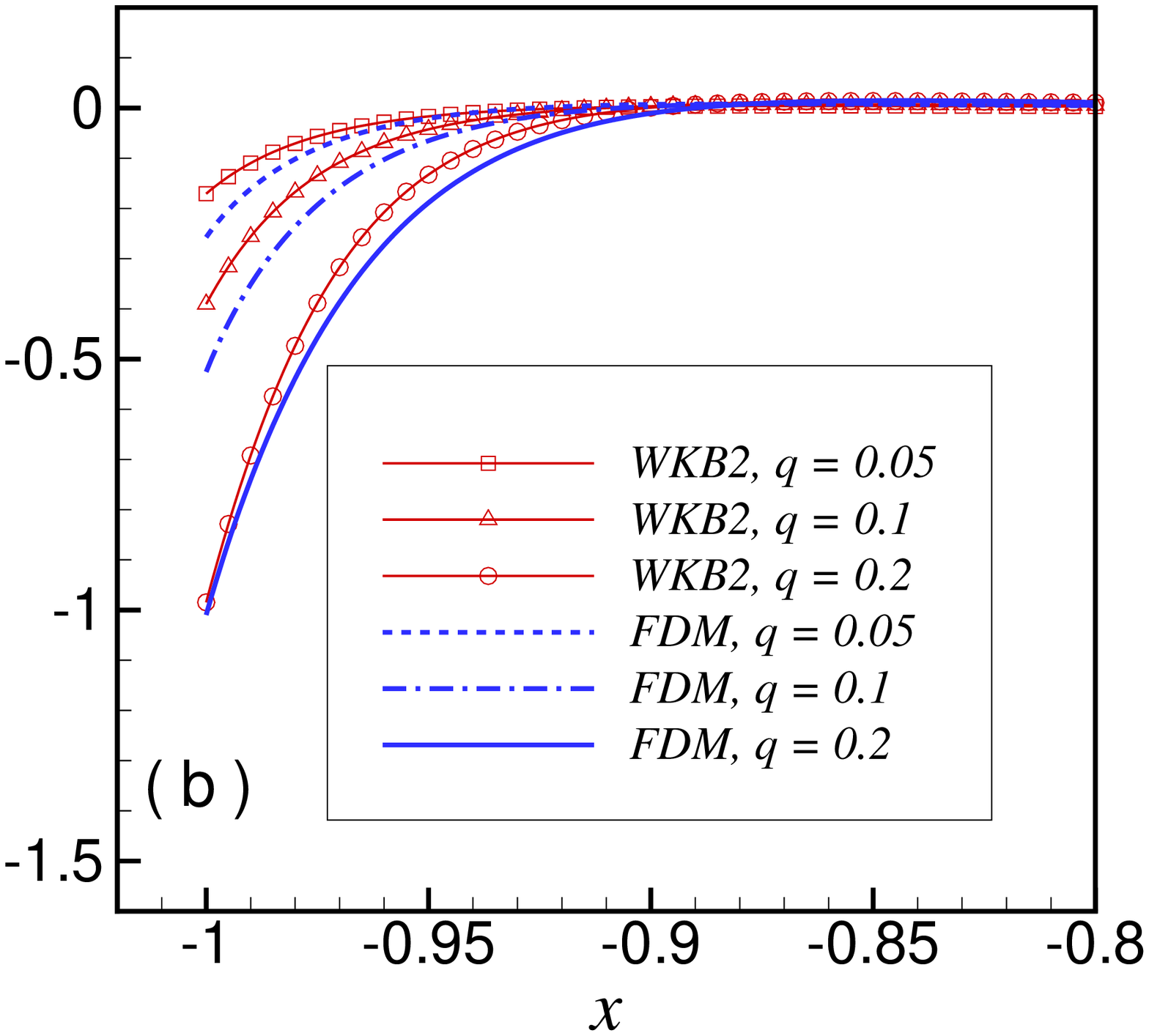}
    \caption{(Color online) Deviation of net charge density from the PNP for different $q$ values at $T=10$ and different approaches. Left: WKB1 vs. FDM; right: WKB2 vs. FDM.}
    \label{fig:dev_rho}
\end{figure*}

We should see that the WKB1 approximation is already inaccurate in the case of small $q$. To make a further comparison discussion, we plot the results of deviating from the PNP by the three approaches in Fig. \ref{fig:dev_rho}. It is obvious the image charge should not be neglected as the charge density near the boundaries will be much smaller when it is present. One can also see that, the WKB2 prediction has been greatly improved from WKB1, and agrees well with the FDM, though the WKB2 uses the asymptotic expression for approximating the self energy.

\subsection{Effect of dielectric ratio}

\begin{figure}[htbp!]
    \centering
    \includegraphics[width=.9\linewidth]{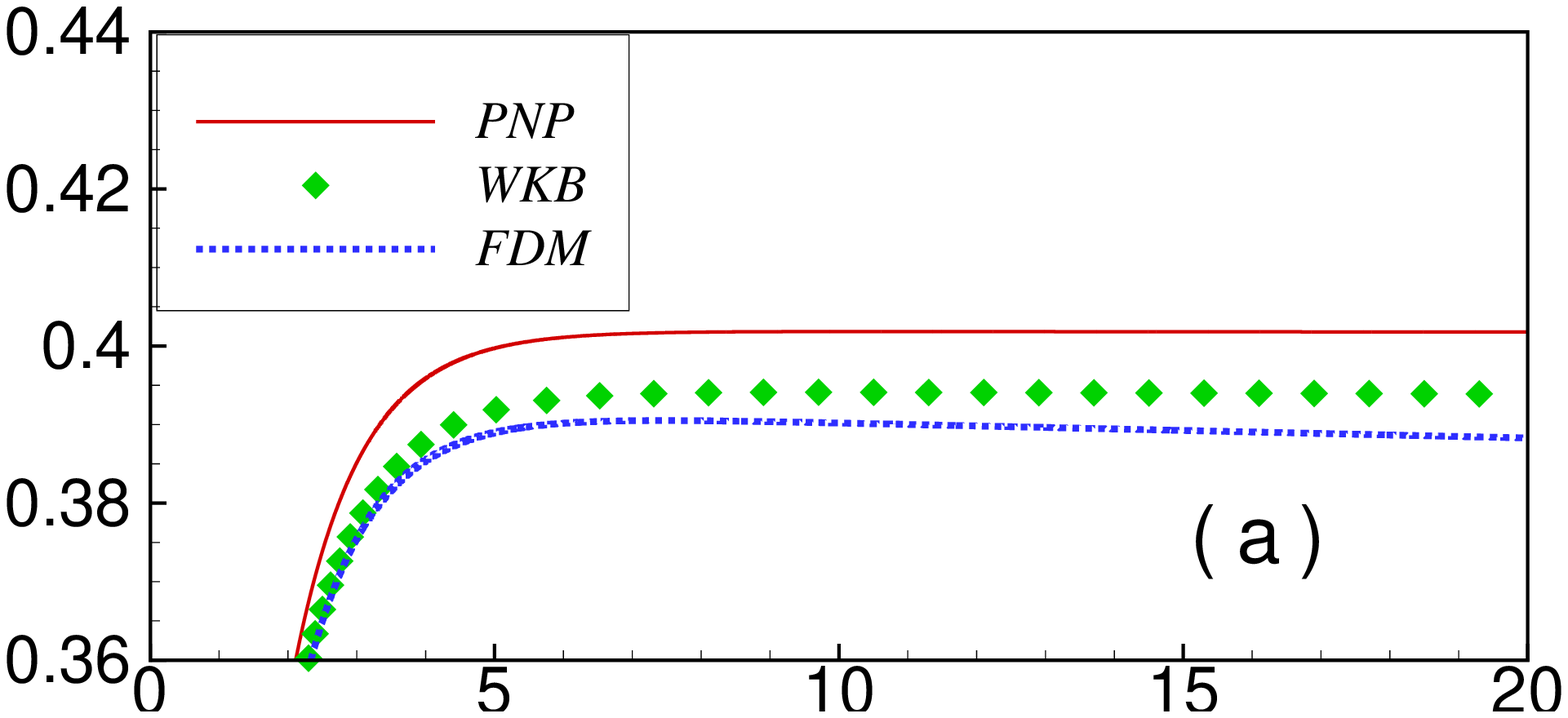}
    \includegraphics[width=.9\linewidth]{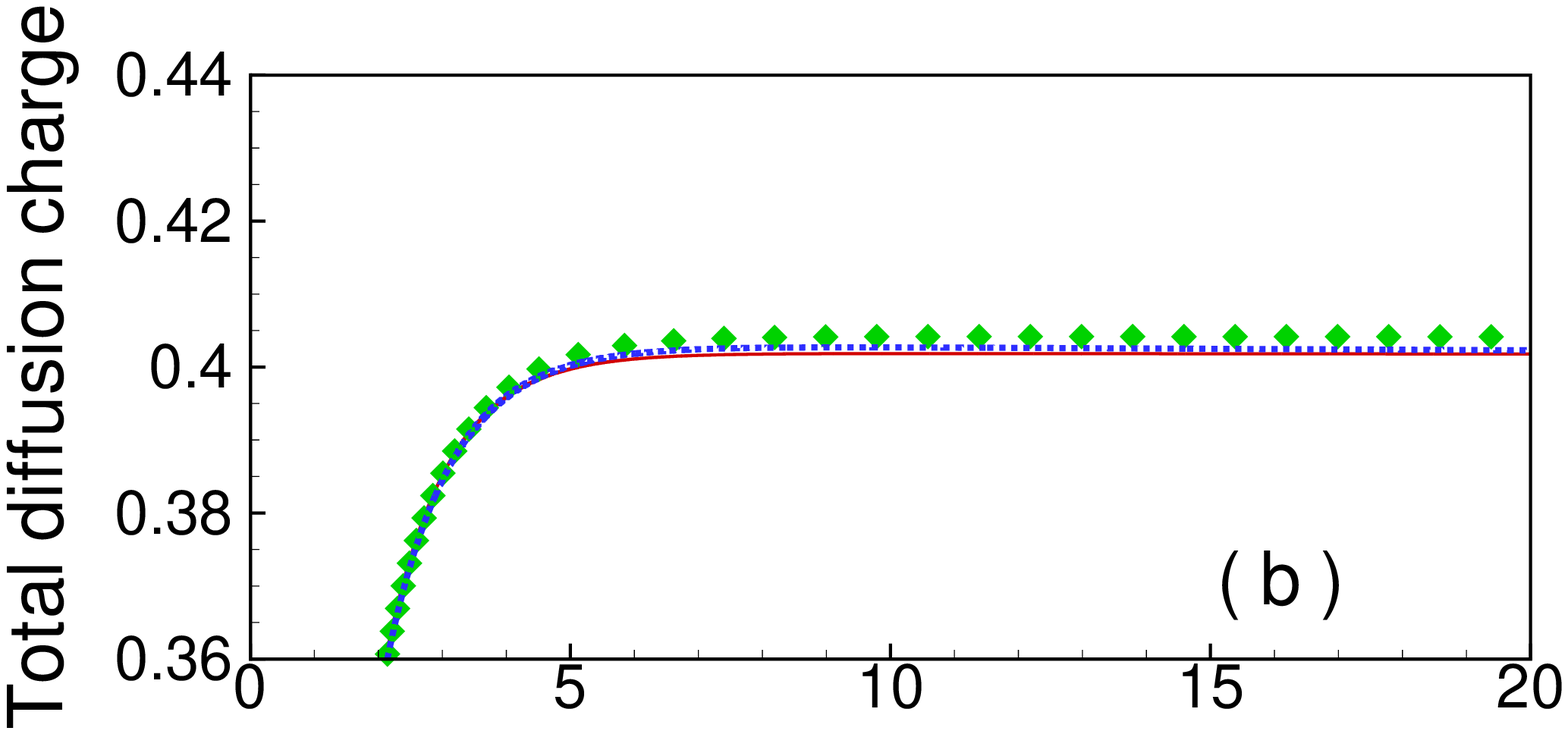}
    \includegraphics[width=.9\linewidth]{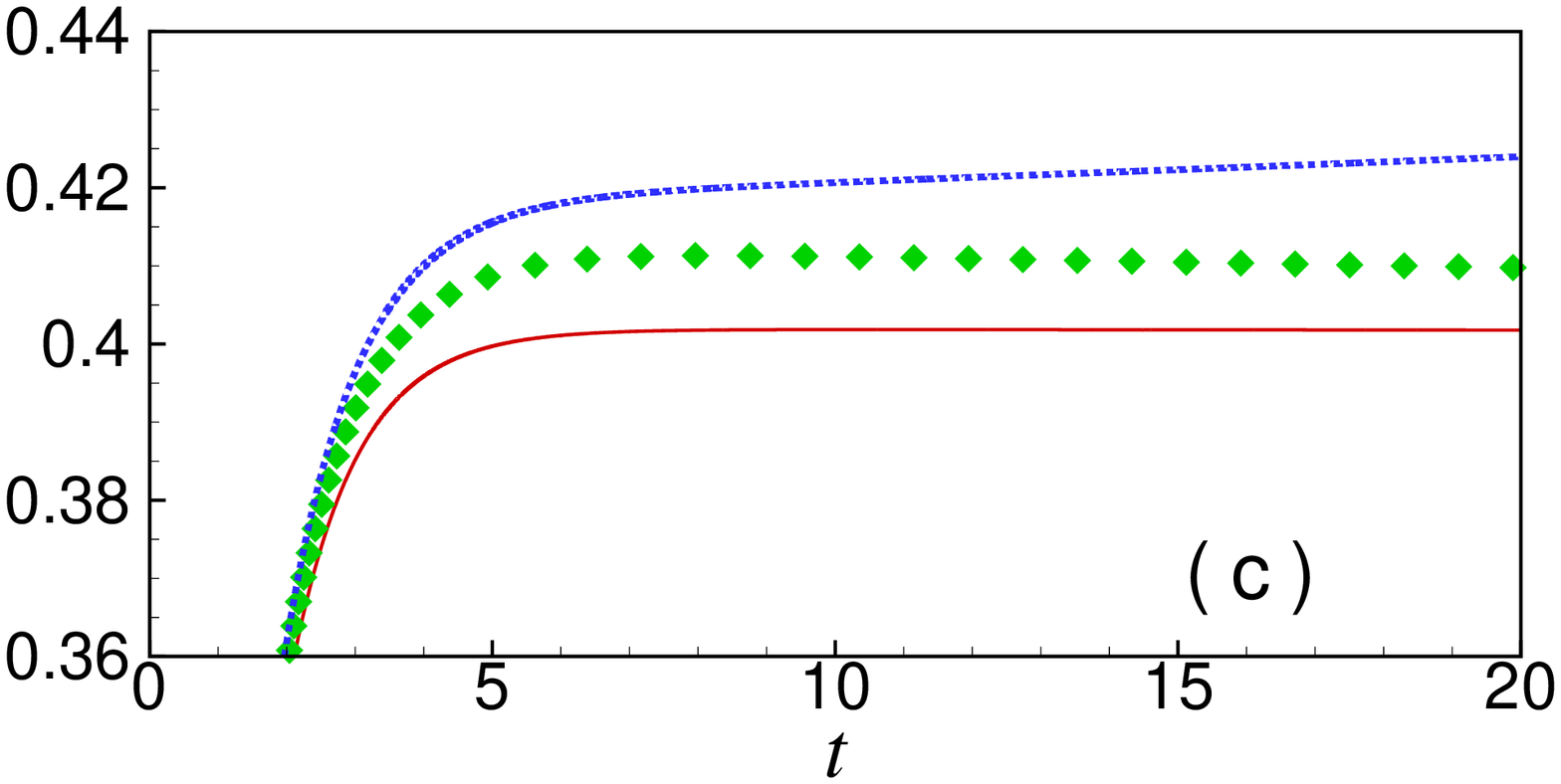}
    \caption{(Color online) Comparison of total diffusion charge as function of $t$. The WKB approximation uses the WKB2. (a) $\varepsilon_B/\varepsilon_W=1/20$; (b) $\varepsilon_B/\varepsilon_W=1$; (c) $\varepsilon_B/\varepsilon_W=20$. }
    \label{fig:d}
\end{figure}

Now, we investigate charge dynamics for different dielectric ratios, $\varepsilon_B/\varepsilon_W=1/20, 1$ and $20.$ We take $q=0.1$ for the modified PNP equations, and calculate the total diffusion charge in the left half of the electrolyte (near the cathode),
\begin{equation}
\rho(t)=\int_{-1}^0 \sum_i z_i c_i(x,t) dx,
\end{equation}
as a time function.  Fig. \ref{fig:d} presents the PNP, WKB2 and FDM results.

When $\varepsilon_B/\varepsilon_W<1$, the image repulsion by FDM and WKB2 reduces  the total diffusion charge, compared to the PNP. It is not difficult to understand that the image charge repulsion leads to a reduction of total charge near the surface. The FDM is slightly smaller than the WKB2 prediction. In the case of $\varepsilon_B/\varepsilon_W=1$ where the polarization effect is gone, the difference between the FDM (or WKB2) and the PNP is minor, and thus the electrostatic correlation between ions is weak.
In the case of $\varepsilon_B/\varepsilon_W>1$, the results illustrate a strong image attraction, which leads to counterion condensation on the surfaces, thus bigger total charges are predicted in comparison to the two previous cases. When time is small, the FDM and WKB2 agree well, but the deviation may be high for large $T$ and large $\varepsilon_B$. This demonstrates the asymptotic approach may be quantitatively inaccurate when the interfacial counterion density has a sharp change.

The FDM results also show that the total charge keeps increasing for large $\varepsilon_B$ (or decreasing for small $\varepsilon_B$) after a long time, demonstrating the existence of larger time scale in the modified PNP model. A possible explanation is the strong image attraction causes a high counterion density at surface where ion-ion correlation leads to a new relaxation time scale. For large $\varepsilon_B$, the solution may blow up at a certain time without including the excluded volume effect.

\section{Concluding remarks}

In summary, we have proposed a modified PNP model by coupling the PNP equation with the generalized DH equation and developed efficient WKB and numerical methods for the self energy. We show by numerical examples both methods are accurate, and the analytical WKB approximation is in good agreement with the difference method.

It has been investigated through simulations that image charges are involved in many many-body phenomena such as charge inversion \cite{Messina:JCP:2002,GXX:JCP:2012} and like-charge attraction \cite{WangRui:JCP:13,Xu:PRE:2013}. We show that the modified model can correctly predict the image-charge effect on dynamics of mobile ions.
However, this work does not pay much attention on the ion-ion correlation, though this effect has been included in the modified PNP model. It is due to that the ion correlation should take into account the size effect of ions, which gives rise to a high difficulty in numerical implementation. Without accounting for the ion size effect, the solution of the modified PNP equation is unstable with the increase of applied surface voltages. The issue of overcoming this challenge is certainly our objective of future work.

\section*{Acknowledgments}
The authors acknowledge the financial support from the Natural Science Foundation of China (Grant Numbers: 11101276, and 91130012), Youth Talents Program by Chinese Organization Department, and the HPC center of SJTU. The authors thank Prof. Bob Eisenberg for careful reading and comments.

%%%% Bibliography  %%%%%%%%%%
%%{unsrt}%{SIAM}% {abbrv} %{nature} %{plain} %{abbrv} %
%\bibliographystyle{elsart-num} %{elsart-num-sort} {plain} %
%\bibliography{C:/Users/zhenli/Dropbox/XuGroup_Share/groupbib}
%\bibliography{D:/Dropbox/XuGroup_Share/groupbib}
%\bibliography{C:/Users/user/Dropbox/Research/XuGroup_Share/groupbib}
%\bibliography{groupbib}

\begin{thebibliography}{68}
\expandafter\ifx\csname natexlab\endcsname\relax\def\natexlab#1{#1}\fi
\expandafter\ifx\csname bibnamefont\endcsname\relax
  \def\bibnamefont#1{#1}\fi
\expandafter\ifx\csname bibfnamefont\endcsname\relax
  \def\bibfnamefont#1{#1}\fi
\expandafter\ifx\csname citenamefont\endcsname\relax
  \def\citenamefont#1{#1}\fi
\expandafter\ifx\csname url\endcsname\relax
  \def\url#1{\texttt{#1}}\fi
\expandafter\ifx\csname urlprefix\endcsname\relax\def\urlprefix{URL }\fi
\providecommand{\bibinfo}[2]{#2}
\providecommand{\eprint}[2][]{\url{#2}}

\bibitem[{\citenamefont{Schoch et~al.}(2008)\citenamefont{Schoch, Han, and
  Renaud}}]{Schoch:RMP:08}
\bibinfo{author}{\bibfnamefont{R.~B.} \bibnamefont{Schoch}},
  \bibinfo{author}{\bibfnamefont{J.}~\bibnamefont{Han}}, \bibnamefont{and}
  \bibinfo{author}{\bibfnamefont{P.}~\bibnamefont{Renaud}},
  \bibinfo{journal}{Rev. Mod. Phys.} \textbf{\bibinfo{volume}{80}},
  \bibinfo{pages}{839} (\bibinfo{year}{2008}).

\bibitem[{\citenamefont{Daiguji}(2010)}]{Daiguji:CSR:10}
\bibinfo{author}{\bibfnamefont{H.}~\bibnamefont{Daiguji}},
  \bibinfo{journal}{Chem. Soc. Rev.} \textbf{\bibinfo{volume}{39}},
  \bibinfo{pages}{901} (\bibinfo{year}{2010}).

\bibitem[{\citenamefont{French et~al.}(2010)\citenamefont{French, Parsegian,
  Podgornik, Rajter, Jagota, Luo, Asthagiri, Chaudhury, Chiang, Granick
  et~al.}}]{FPP+:RMP:2010}
\bibinfo{author}{\bibfnamefont{R.~H.} \bibnamefont{French}},
  \bibinfo{author}{\bibfnamefont{V.~A.} \bibnamefont{Parsegian}},
  \bibinfo{author}{\bibfnamefont{R.}~\bibnamefont{Podgornik}},
  \bibinfo{author}{\bibfnamefont{R.~F.} \bibnamefont{Rajter}},
  \bibinfo{author}{\bibfnamefont{A.}~\bibnamefont{Jagota}},
  \bibinfo{author}{\bibfnamefont{J.}~\bibnamefont{Luo}},
  \bibinfo{author}{\bibfnamefont{D.}~\bibnamefont{Asthagiri}},
  \bibinfo{author}{\bibfnamefont{M.~K.} \bibnamefont{Chaudhury}},
  \bibinfo{author}{\bibfnamefont{Y.-M.} \bibnamefont{Chiang}},
  \bibinfo{author}{\bibfnamefont{S.}~\bibnamefont{Granick}},
  \bibnamefont{et~al.}, \bibinfo{journal}{Rev. Mod. Phys.}
  \textbf{\bibinfo{volume}{82}}, \bibinfo{pages}{1887} (\bibinfo{year}{2010}).

\bibitem[{\citenamefont{Levin}(2002)}]{Levin:RPP:2002}
\bibinfo{author}{\bibfnamefont{Y.}~\bibnamefont{Levin}}, \bibinfo{journal}{Rep.
  Prog. Phys.} \textbf{\bibinfo{volume}{65}}, \bibinfo{pages}{1577}
  (\bibinfo{year}{2002}).

\bibitem[{\citenamefont{Walker et~al.}(2011)\citenamefont{Walker, Kowalczyk,
  {de la Cruz}, and Grzybowski}}]{WK+:N:2011}
\bibinfo{author}{\bibfnamefont{D.~A.} \bibnamefont{Walker}},
  \bibinfo{author}{\bibfnamefont{B.}~\bibnamefont{Kowalczyk}},
  \bibinfo{author}{\bibfnamefont{M.~O.} \bibnamefont{{de la Cruz}}},
  \bibnamefont{and} \bibinfo{author}{\bibfnamefont{B.~A.}
  \bibnamefont{Grzybowski}}, \bibinfo{journal}{Nanoscale}
  \textbf{\bibinfo{volume}{3}}, \bibinfo{pages}{1316} (\bibinfo{year}{2011}).

\bibitem[{\citenamefont{Messina}(2009)}]{Messina:JPCM:2009}
\bibinfo{author}{\bibfnamefont{R.}~\bibnamefont{Messina}}, \bibinfo{journal}{J.
  Phys. Condens. Matter} \textbf{\bibinfo{volume}{21}}, \bibinfo{pages}{113102}
  (\bibinfo{year}{2009}).

\bibitem[{\citenamefont{Wan et~al.}(2014)\citenamefont{Wan, Xu, Liao, Liu, and
  Sheng}}]{wan2014:PRX}
\bibinfo{author}{\bibfnamefont{L.}~\bibnamefont{Wan}},
  \bibinfo{author}{\bibfnamefont{S.}~\bibnamefont{Xu}},
  \bibinfo{author}{\bibfnamefont{M.}~\bibnamefont{Liao}},
  \bibinfo{author}{\bibfnamefont{C.}~\bibnamefont{Liu}}, \bibnamefont{and}
  \bibinfo{author}{\bibfnamefont{P.}~\bibnamefont{Sheng}},
  \bibinfo{journal}{Phys. Rev. X} \textbf{\bibinfo{volume}{4}},
  \bibinfo{pages}{011042} (\bibinfo{year}{2014}).

\bibitem[{\citenamefont{Messina}(2002)}]{Messina:JCP:2002}
\bibinfo{author}{\bibfnamefont{R.}~\bibnamefont{Messina}}, \bibinfo{journal}{J.
  Chem. Phys.} \textbf{\bibinfo{volume}{117}}, \bibinfo{pages}{11062}
  (\bibinfo{year}{2002}).

\bibitem[{\citenamefont{Hatlo and Lue}(2008)}]{HL:SM:2008}
\bibinfo{author}{\bibfnamefont{M.~M.} \bibnamefont{Hatlo}} \bibnamefont{and}
  \bibinfo{author}{\bibfnamefont{L.}~\bibnamefont{Lue}}, \bibinfo{journal}{Soft
  Matter} \textbf{\bibinfo{volume}{4}}, \bibinfo{pages}{1582}
  (\bibinfo{year}{2008}).

\bibitem[{\citenamefont{Jho et~al.}(2008)\citenamefont{Jho,
  Kandu\ifmmode~\check{c}\else \v{c}\fi{}, Naji, Podgornik, Kim, and
  Pincus}}]{JKN+:PRL:2008}
\bibinfo{author}{\bibfnamefont{Y.~S.} \bibnamefont{Jho}},
  \bibinfo{author}{\bibfnamefont{M.}~\bibnamefont{Kandu\ifmmode~\check{c}\else
  \v{c}\fi{}}}, \bibinfo{author}{\bibfnamefont{A.}~\bibnamefont{Naji}},
  \bibinfo{author}{\bibfnamefont{R.}~\bibnamefont{Podgornik}},
  \bibinfo{author}{\bibfnamefont{M.~W.} \bibnamefont{Kim}}, \bibnamefont{and}
  \bibinfo{author}{\bibfnamefont{P.~A.} \bibnamefont{Pincus}},
  \bibinfo{journal}{Phys. Rev. Lett.} \textbf{\bibinfo{volume}{101}},
  \bibinfo{pages}{188101} (\bibinfo{year}{2008}).

\bibitem[{\citenamefont{Wang and Ma}(2010)}]{WM:JPCB:2010}
\bibinfo{author}{\bibfnamefont{Z.~Y.} \bibnamefont{Wang}} \bibnamefont{and}
  \bibinfo{author}{\bibfnamefont{Y.~Q.} \bibnamefont{Ma}}, \bibinfo{journal}{J.
  Phys. Chem. B} \textbf{\bibinfo{volume}{114}}, \bibinfo{pages}{13386}
  (\bibinfo{year}{2010}).

\bibitem[{\citenamefont{Bakhshandeh et~al.}(2011)\citenamefont{Bakhshandeh, dos
  Santos, and Levin}}]{Bakhshandeh:PRL:2011}
\bibinfo{author}{\bibfnamefont{A.}~\bibnamefont{Bakhshandeh}},
  \bibinfo{author}{\bibfnamefont{A.~P.} \bibnamefont{dos Santos}},
  \bibnamefont{and} \bibinfo{author}{\bibfnamefont{Y.}~\bibnamefont{Levin}},
  \bibinfo{journal}{Phys. Rev. Lett.} \textbf{\bibinfo{volume}{107}},
  \bibinfo{pages}{107801} (\bibinfo{year}{2011}).

\bibitem[{\citenamefont{Diehl et~al.}(2012)\citenamefont{Diehl, {dos Santos},
  and Levin}}]{DL:JPCM:2012a}
\bibinfo{author}{\bibfnamefont{A.}~\bibnamefont{Diehl}},
  \bibinfo{author}{\bibfnamefont{A.~P.} \bibnamefont{{dos Santos}}},
  \bibnamefont{and} \bibinfo{author}{\bibfnamefont{Y.}~\bibnamefont{Levin}},
  \bibinfo{journal}{J. Phys.: Condens. Matter} \textbf{\bibinfo{volume}{24}},
  \bibinfo{pages}{284115} (\bibinfo{year}{2012}).

\bibitem[{\citenamefont{Gan et~al.}(2012)\citenamefont{Gan, Xing, and
  Xu}}]{GXX:JCP:2012}
\bibinfo{author}{\bibfnamefont{Z.}~\bibnamefont{Gan}},
  \bibinfo{author}{\bibfnamefont{X.}~\bibnamefont{Xing}}, \bibnamefont{and}
  \bibinfo{author}{\bibfnamefont{Z.}~\bibnamefont{Xu}}, \bibinfo{journal}{J.
  Chem. Phys.} \textbf{\bibinfo{volume}{137}}, \bibinfo{pages}{034708}
  (\bibinfo{year}{2012}).

\bibitem[{\citenamefont{Xu}(2013)}]{Xu:PRE:2013}
\bibinfo{author}{\bibfnamefont{Z.}~\bibnamefont{Xu}}, \bibinfo{journal}{Phys.
  Rev. E} \textbf{\bibinfo{volume}{87}}, \bibinfo{pages}{013307}
  (\bibinfo{year}{2013}).

\bibitem[{\citenamefont{Wang and Wang}(2013)}]{WangRui:JCP:13}
\bibinfo{author}{\bibfnamefont{R.}~\bibnamefont{Wang}} \bibnamefont{and}
  \bibinfo{author}{\bibfnamefont{Z.-G.} \bibnamefont{Wang}},
  \bibinfo{journal}{J. Chem. Phys.} \textbf{\bibinfo{volume}{139}},
  \bibinfo{pages}{124702} (\bibinfo{year}{2013}).

\bibitem[{\citenamefont{Zwanikken and {de la Cruz}}(2013)}]{ZO:PNAS:13}
\bibinfo{author}{\bibfnamefont{J.~W.} \bibnamefont{Zwanikken}}
  \bibnamefont{and} \bibinfo{author}{\bibfnamefont{M.~O.} \bibnamefont{{de la
  Cruz}}}, \bibinfo{journal}{Proc. Nat. Acad. Sci. USA}
  \textbf{\bibinfo{volume}{110}}, \bibinfo{pages}{5301} (\bibinfo{year}{2013}).

\bibitem[{\citenamefont{Chen and Eisenberg}(1993)}]{CE:BJ:93}
\bibinfo{author}{\bibfnamefont{D.}~\bibnamefont{Chen}} \bibnamefont{and}
  \bibinfo{author}{\bibfnamefont{R.}~\bibnamefont{Eisenberg}},
  \bibinfo{journal}{Biophys. J.} \textbf{\bibinfo{volume}{64}},
  \bibinfo{pages}{1405} (\bibinfo{year}{1993}).

\bibitem[{\citenamefont{Eisenberg}(1996)}]{E:JMB:96}
\bibinfo{author}{\bibfnamefont{R.~S.} \bibnamefont{Eisenberg}},
  \bibinfo{journal}{J. Membrane Biol.} \textbf{\bibinfo{volume}{150}},
  \bibinfo{pages}{1} (\bibinfo{year}{1996}).

\bibitem[{\citenamefont{Eisenberg}(1999)}]{E:JMB:99}
\bibinfo{author}{\bibfnamefont{R.~S.} \bibnamefont{Eisenberg}},
  \bibinfo{journal}{J. Membrane Biol.} \textbf{\bibinfo{volume}{171}},
  \bibinfo{pages}{1} (\bibinfo{year}{1999}).

\bibitem[{\citenamefont{Hollerbach et~al.}(2001)\citenamefont{Hollerbach, Chen,
  and Eisenberg}}]{HCE:JSC:01}
\bibinfo{author}{\bibfnamefont{U.}~\bibnamefont{Hollerbach}},
  \bibinfo{author}{\bibfnamefont{D.-P.} \bibnamefont{Chen}}, \bibnamefont{and}
  \bibinfo{author}{\bibfnamefont{R.}~\bibnamefont{Eisenberg}},
  \bibinfo{journal}{J. Sci. Comput.} \textbf{\bibinfo{volume}{16}},
  \bibinfo{pages}{373} (\bibinfo{year}{2001}).

\bibitem[{\citenamefont{Daiguji et~al.}(2005)\citenamefont{Daiguji, Oka, and
  Shirono}}]{daiguji:NL:2005}
\bibinfo{author}{\bibfnamefont{H.}~\bibnamefont{Daiguji}},
  \bibinfo{author}{\bibfnamefont{Y.}~\bibnamefont{Oka}}, \bibnamefont{and}
  \bibinfo{author}{\bibfnamefont{K.}~\bibnamefont{Shirono}},
  \bibinfo{journal}{Nano Letters} \textbf{\bibinfo{volume}{5}},
  \bibinfo{pages}{2274} (\bibinfo{year}{2005}).

\bibitem[{\citenamefont{Markowich et~al.}(1990)\citenamefont{Markowich,
  Ringhofer, and Schimeiser}}]{MRS:S:90}
\bibinfo{author}{\bibfnamefont{P.}~\bibnamefont{Markowich}},
  \bibinfo{author}{\bibfnamefont{C.}~\bibnamefont{Ringhofer}},
  \bibnamefont{and}
  \bibinfo{author}{\bibfnamefont{C.}~\bibnamefont{Schimeiser}},
  \emph{\bibinfo{title}{Semiconductor}} (\bibinfo{publisher}{Springer},
  \bibinfo{year}{1990}).

\bibitem[{\citenamefont{Bazant et~al.}(2004)\citenamefont{Bazant, Thornton, and
  Ajdari}}]{BTA:PRE:04}
\bibinfo{author}{\bibfnamefont{M.~Z.} \bibnamefont{Bazant}},
  \bibinfo{author}{\bibfnamefont{K.}~\bibnamefont{Thornton}}, \bibnamefont{and}
  \bibinfo{author}{\bibfnamefont{A.}~\bibnamefont{Ajdari}},
  \bibinfo{journal}{Physical review E} \textbf{\bibinfo{volume}{70}},
  \bibinfo{pages}{021506} (\bibinfo{year}{2004}).

\bibitem[{\citenamefont{Squires and Bazant}(2004)}]{SB:JFM:04}
\bibinfo{author}{\bibfnamefont{T.~M.} \bibnamefont{Squires}} \bibnamefont{and}
  \bibinfo{author}{\bibfnamefont{M.~Z.} \bibnamefont{Bazant}},
  \bibinfo{journal}{Journal of Fluid Mechanics} \textbf{\bibinfo{volume}{509}},
  \bibinfo{pages}{217} (\bibinfo{year}{2004}).

\bibitem[{\citenamefont{Singer and Norbury}(2009)}]{SN:SIAM:09}
\bibinfo{author}{\bibfnamefont{A.}~\bibnamefont{Singer}} \bibnamefont{and}
  \bibinfo{author}{\bibfnamefont{J.}~\bibnamefont{Norbury}},
  \bibinfo{journal}{SIAM J. Appl. Math.} \textbf{\bibinfo{volume}{70}},
  \bibinfo{pages}{949} (\bibinfo{year}{2009}).

\bibitem[{\citenamefont{Kurnikova et~al.}(1999)\citenamefont{Kurnikova,
  Coalson, Graf, and Nitzan}}]{KCG:BJ:99}
\bibinfo{author}{\bibfnamefont{M.~G.} \bibnamefont{Kurnikova}},
  \bibinfo{author}{\bibfnamefont{R.~D.} \bibnamefont{Coalson}},
  \bibinfo{author}{\bibfnamefont{P.}~\bibnamefont{Graf}}, \bibnamefont{and}
  \bibinfo{author}{\bibfnamefont{A.}~\bibnamefont{Nitzan}},
  \bibinfo{journal}{Biophys. J.} \textbf{\bibinfo{volume}{76}},
  \bibinfo{pages}{642} (\bibinfo{year}{1999}).

\bibitem[{\citenamefont{Cardenas et~al.}(2000)\citenamefont{Cardenas, Coalson,
  and Kurnikova}}]{CCK:BJ:00}
\bibinfo{author}{\bibfnamefont{A.~E.} \bibnamefont{Cardenas}},
  \bibinfo{author}{\bibfnamefont{R.~D.} \bibnamefont{Coalson}},
  \bibnamefont{and} \bibinfo{author}{\bibfnamefont{M.~G.}
  \bibnamefont{Kurnikova}}, \bibinfo{journal}{Biophys. J.}
  \textbf{\bibinfo{volume}{79}}, \bibinfo{pages}{80} (\bibinfo{year}{2000}).

\bibitem[{\citenamefont{Zheng et~al.}(2011)\citenamefont{Zheng, Chen, and
  Wei}}]{ZCW:JCP:11}
\bibinfo{author}{\bibfnamefont{Q.}~\bibnamefont{Zheng}},
  \bibinfo{author}{\bibfnamefont{D.}~\bibnamefont{Chen}}, \bibnamefont{and}
  \bibinfo{author}{\bibfnamefont{G.-W.} \bibnamefont{Wei}},
  \bibinfo{journal}{J. Comput. Phys.} \textbf{\bibinfo{volume}{230}},
  \bibinfo{pages}{5239} (\bibinfo{year}{2011}).

\bibitem[{\citenamefont{Wei et~al.}(2012)\citenamefont{Wei, Zheng, Chen, and
  Xia}}]{WZCX:SIAM:12}
\bibinfo{author}{\bibfnamefont{G.-W.} \bibnamefont{Wei}},
  \bibinfo{author}{\bibfnamefont{Q.}~\bibnamefont{Zheng}},
  \bibinfo{author}{\bibfnamefont{Z.}~\bibnamefont{Chen}}, \bibnamefont{and}
  \bibinfo{author}{\bibfnamefont{K.}~\bibnamefont{Xia}}, \bibinfo{journal}{SIAM
  Review} \textbf{\bibinfo{volume}{54}}, \bibinfo{pages}{699}
  (\bibinfo{year}{2012}).

\bibitem[{\citenamefont{Flavell et~al.}(2014)\citenamefont{Flavell, Machen,
  Eisenberg, Kabre, Liu, and Li}}]{Flavell:JCE:14}
\bibinfo{author}{\bibfnamefont{A.}~\bibnamefont{Flavell}},
  \bibinfo{author}{\bibfnamefont{M.}~\bibnamefont{Machen}},
  \bibinfo{author}{\bibfnamefont{B.}~\bibnamefont{Eisenberg}},
  \bibinfo{author}{\bibfnamefont{J.}~\bibnamefont{Kabre}},
  \bibinfo{author}{\bibfnamefont{C.}~\bibnamefont{Liu}}, \bibnamefont{and}
  \bibinfo{author}{\bibfnamefont{X.}~\bibnamefont{Li}}, \bibinfo{journal}{J.
  Comput. Electronics} \textbf{\bibinfo{volume}{13}}, \bibinfo{pages}{235}
  (\bibinfo{year}{2014}).

\bibitem[{\citenamefont{Dyrka et~al.}(2013)\citenamefont{Dyrka, Bartuzel, and
  Kotulska}}]{dyrka2013Proteins}
\bibinfo{author}{\bibfnamefont{W.}~\bibnamefont{Dyrka}},
  \bibinfo{author}{\bibfnamefont{M.~M.} \bibnamefont{Bartuzel}},
  \bibnamefont{and} \bibinfo{author}{\bibfnamefont{M.}~\bibnamefont{Kotulska}},
  \bibinfo{journal}{Proteins: Structure, Function, and Bioinformatics}
  \textbf{\bibinfo{volume}{81}}, \bibinfo{pages}{1802} (\bibinfo{year}{2013}).

\bibitem[{\citenamefont{Lockett et~al.}(2008)\citenamefont{Lockett, Sedev,
  Ralston, Horne, and Rodopoulos}}]{LSR:JPC:08}
\bibinfo{author}{\bibfnamefont{V.}~\bibnamefont{Lockett}},
  \bibinfo{author}{\bibfnamefont{R.}~\bibnamefont{Sedev}},
  \bibinfo{author}{\bibfnamefont{J.}~\bibnamefont{Ralston}},
  \bibinfo{author}{\bibfnamefont{M.}~\bibnamefont{Horne}}, \bibnamefont{and}
  \bibinfo{author}{\bibfnamefont{T.}~\bibnamefont{Rodopoulos}},
  \bibinfo{journal}{The J. Phys. Chem. C} \textbf{\bibinfo{volume}{112}},
  \bibinfo{pages}{7486} (\bibinfo{year}{2008}).

\bibitem[{\citenamefont{Mamonov et~al.}(2003)\citenamefont{Mamonov, Coalson,
  Nitzan, and Kurnikova}}]{MRA:BJ:03}
\bibinfo{author}{\bibfnamefont{A.~B.} \bibnamefont{Mamonov}},
  \bibinfo{author}{\bibfnamefont{R.~D.} \bibnamefont{Coalson}},
  \bibinfo{author}{\bibfnamefont{A.}~\bibnamefont{Nitzan}}, \bibnamefont{and}
  \bibinfo{author}{\bibfnamefont{M.~G.} \bibnamefont{Kurnikova}},
  \bibinfo{journal}{Biophys. J.} \textbf{\bibinfo{volume}{84}},
  \bibinfo{pages}{3646 } (\bibinfo{year}{2003}).

\bibitem[{\citenamefont{Busath et~al.}(1998)\citenamefont{Busath, Thulin,
  Hendershot, Phillips, Maughan, Cole, Bingham, Morrison, Baird, Hendershot
  et~al.}}]{BTH:BJ:98}
\bibinfo{author}{\bibfnamefont{D.~D.} \bibnamefont{Busath}},
  \bibinfo{author}{\bibfnamefont{C.~D.} \bibnamefont{Thulin}},
  \bibinfo{author}{\bibfnamefont{R.~W.} \bibnamefont{Hendershot}},
  \bibinfo{author}{\bibfnamefont{L.~R.} \bibnamefont{Phillips}},
  \bibinfo{author}{\bibfnamefont{P.}~\bibnamefont{Maughan}},
  \bibinfo{author}{\bibfnamefont{C.~D.} \bibnamefont{Cole}},
  \bibinfo{author}{\bibfnamefont{N.~C.} \bibnamefont{Bingham}},
  \bibinfo{author}{\bibfnamefont{S.}~\bibnamefont{Morrison}},
  \bibinfo{author}{\bibfnamefont{L.~C.} \bibnamefont{Baird}},
  \bibinfo{author}{\bibfnamefont{R.~J.} \bibnamefont{Hendershot}},
  \bibnamefont{et~al.}, \bibinfo{journal}{Biophys. J.}
  \textbf{\bibinfo{volume}{75}}, \bibinfo{pages}{2830} (\bibinfo{year}{1998}).

\bibitem[{\citenamefont{K{\'e}kicheff and Spalla}(1995)}]{KS:PRL:95}
\bibinfo{author}{\bibfnamefont{P.}~\bibnamefont{K{\'e}kicheff}}
  \bibnamefont{and} \bibinfo{author}{\bibfnamefont{O.}~\bibnamefont{Spalla}},
  \bibinfo{journal}{Phys. Rev. Lett.} \textbf{\bibinfo{volume}{75}},
  \bibinfo{pages}{1851} (\bibinfo{year}{1995}).

\bibitem[{\citenamefont{Corry et~al.}(2000)\citenamefont{Corry, Kuyucak, and
  Chung}}]{CKC:BJ:00}
\bibinfo{author}{\bibfnamefont{B.}~\bibnamefont{Corry}},
  \bibinfo{author}{\bibfnamefont{S.}~\bibnamefont{Kuyucak}}, \bibnamefont{and}
  \bibinfo{author}{\bibfnamefont{S.-H.} \bibnamefont{Chung}},
  \bibinfo{journal}{Biophys. J.} \textbf{\bibinfo{volume}{78}},
  \bibinfo{pages}{2364} (\bibinfo{year}{2000}).

\bibitem[{\citenamefont{Storey et~al.}(2008)\citenamefont{Storey, Edwards,
  Kilic, and Bazant}}]{storey2008pre}
\bibinfo{author}{\bibfnamefont{B.~D.} \bibnamefont{Storey}},
  \bibinfo{author}{\bibfnamefont{L.~R.} \bibnamefont{Edwards}},
  \bibinfo{author}{\bibfnamefont{M.~S.} \bibnamefont{Kilic}}, \bibnamefont{and}
  \bibinfo{author}{\bibfnamefont{M.~Z.} \bibnamefont{Bazant}},
  \bibinfo{journal}{Physical Review E} \textbf{\bibinfo{volume}{77}},
  \bibinfo{pages}{036317} (\bibinfo{year}{2008}).

\bibitem[{\citenamefont{Kilic et~al.}(2007)\citenamefont{Kilic, Bazant, and
  Ajdari}}]{KBA:PRE:07}
\bibinfo{author}{\bibfnamefont{M.~S.} \bibnamefont{Kilic}},
  \bibinfo{author}{\bibfnamefont{M.~Z.} \bibnamefont{Bazant}},
  \bibnamefont{and} \bibinfo{author}{\bibfnamefont{A.}~\bibnamefont{Ajdari}},
  \bibinfo{journal}{Physical Review E} \textbf{\bibinfo{volume}{75}},
  \bibinfo{pages}{021503} (\bibinfo{year}{2007}).

\bibitem[{\citenamefont{Horng et~al.}(2012)\citenamefont{Horng, Lin, Liu, and
  Eisenberg}}]{HTLE:JPC:12}
\bibinfo{author}{\bibfnamefont{T.-L.} \bibnamefont{Horng}},
  \bibinfo{author}{\bibfnamefont{T.-C.} \bibnamefont{Lin}},
  \bibinfo{author}{\bibfnamefont{C.}~\bibnamefont{Liu}}, \bibnamefont{and}
  \bibinfo{author}{\bibfnamefont{B.}~\bibnamefont{Eisenberg}},
  \bibinfo{journal}{J. Phys. Chem. B} \textbf{\bibinfo{volume}{116}},
  \bibinfo{pages}{11422} (\bibinfo{year}{2012}).

\bibitem[{\citenamefont{Lu and Zhou}(2011)}]{LZ:BJ:2011}
\bibinfo{author}{\bibfnamefont{B.}~\bibnamefont{Lu}} \bibnamefont{and}
  \bibinfo{author}{\bibfnamefont{Y.}~\bibnamefont{Zhou}},
  \bibinfo{journal}{Biophys. J.} \textbf{\bibinfo{volume}{100}},
  \bibinfo{pages}{2475} (\bibinfo{year}{2011}).

\bibitem[{\citenamefont{Eisenberg et~al.}(2010)\citenamefont{Eisenberg, Hyon,
  and Liu}}]{EHL:JCP:2010}
\bibinfo{author}{\bibfnamefont{B.}~\bibnamefont{Eisenberg}},
  \bibinfo{author}{\bibfnamefont{Y.}~\bibnamefont{Hyon}}, \bibnamefont{and}
  \bibinfo{author}{\bibfnamefont{C.}~\bibnamefont{Liu}}, \bibinfo{journal}{J.
  Chem. Phys.} \textbf{\bibinfo{volume}{133}}, \bibinfo{pages}{104104}
  (\bibinfo{year}{2010}).

\bibitem[{\citenamefont{Liu et~al.}(2012)\citenamefont{Liu, Tu, and
  Zhang}}]{liu2012jdde}
\bibinfo{author}{\bibfnamefont{W.}~\bibnamefont{Liu}},
  \bibinfo{author}{\bibfnamefont{X.}~\bibnamefont{Tu}}, \bibnamefont{and}
  \bibinfo{author}{\bibfnamefont{M.}~\bibnamefont{Zhang}}, \bibinfo{journal}{J.
  Dynamics Diff. Equ.} \textbf{\bibinfo{volume}{24}}, \bibinfo{pages}{985}
  (\bibinfo{year}{2012}).

\bibitem[{\citenamefont{Nadler et~al.}(2003)\citenamefont{Nadler, Hollerbach,
  and Eisenberg}}]{nadler2003pre}
\bibinfo{author}{\bibfnamefont{B.}~\bibnamefont{Nadler}},
  \bibinfo{author}{\bibfnamefont{U.}~\bibnamefont{Hollerbach}},
  \bibnamefont{and} \bibinfo{author}{\bibfnamefont{R.~S.}
  \bibnamefont{Eisenberg}}, \bibinfo{journal}{Phys. Rev. E}
  \textbf{\bibinfo{volume}{68}}, \bibinfo{pages}{021905}
  (\bibinfo{year}{2003}).

\bibitem[{\citenamefont{Corry et~al.}(2003)\citenamefont{Corry, Kuyucak, and
  Chung}}]{CKC:BJ:03}
\bibinfo{author}{\bibfnamefont{B.}~\bibnamefont{Corry}},
  \bibinfo{author}{\bibfnamefont{S.}~\bibnamefont{Kuyucak}}, \bibnamefont{and}
  \bibinfo{author}{\bibfnamefont{S.-H.} \bibnamefont{Chung}},
  \bibinfo{journal}{Biophys. J.} \textbf{\bibinfo{volume}{84}},
  \bibinfo{pages}{3594} (\bibinfo{year}{2003}).

\bibitem[{\citenamefont{Graf et~al.}(2004)\citenamefont{Graf, Kurnikova,
  Coalson, and Nitzan}}]{GKCN:JPC:04}
\bibinfo{author}{\bibfnamefont{P.}~\bibnamefont{Graf}},
  \bibinfo{author}{\bibfnamefont{M.~G.} \bibnamefont{Kurnikova}},
  \bibinfo{author}{\bibfnamefont{R.~D.} \bibnamefont{Coalson}},
  \bibnamefont{and} \bibinfo{author}{\bibfnamefont{A.}~\bibnamefont{Nitzan}},
  \bibinfo{journal}{J. Phys. Chem. B} \textbf{\bibinfo{volume}{108}},
  \bibinfo{pages}{2006} (\bibinfo{year}{2004}).

\bibitem[{\citenamefont{Hyon et~al.}(2011)\citenamefont{Hyon, Eisenberg, and
  Liu}}]{HEL:CMS:11}
\bibinfo{author}{\bibfnamefont{Y.}~\bibnamefont{Hyon}},
  \bibinfo{author}{\bibfnamefont{B.}~\bibnamefont{Eisenberg}},
  \bibnamefont{and} \bibinfo{author}{\bibfnamefont{C.}~\bibnamefont{Liu}},
  \bibinfo{journal}{Commun. Math. Sci} \textbf{\bibinfo{volume}{9}},
  \bibinfo{pages}{459} (\bibinfo{year}{2011}).

\bibitem[{\citenamefont{Luo et~al.}(2006)\citenamefont{Luo, Malkova, Yoon,
  Schultz, Lin, Meron, Benjamin, Vanýsek, and Schlossman}}]{Luo06}
\bibinfo{author}{\bibfnamefont{G.}~\bibnamefont{Luo}},
  \bibinfo{author}{\bibfnamefont{S.}~\bibnamefont{Malkova}},
  \bibinfo{author}{\bibfnamefont{J.}~\bibnamefont{Yoon}},
  \bibinfo{author}{\bibfnamefont{D.~G.} \bibnamefont{Schultz}},
  \bibinfo{author}{\bibfnamefont{B.}~\bibnamefont{Lin}},
  \bibinfo{author}{\bibfnamefont{M.}~\bibnamefont{Meron}},
  \bibinfo{author}{\bibfnamefont{I.}~\bibnamefont{Benjamin}},
  \bibinfo{author}{\bibfnamefont{P.}~\bibnamefont{Vanýsek}}, \bibnamefont{and}
  \bibinfo{author}{\bibfnamefont{M.~L.} \bibnamefont{Schlossman}},
  \bibinfo{journal}{Science} \textbf{\bibinfo{volume}{311}},
  \bibinfo{pages}{216} (\bibinfo{year}{2006}).

\bibitem[{\citenamefont{Podgornik}(1989)}]{podgornik1989jcp}
\bibinfo{author}{\bibfnamefont{R.}~\bibnamefont{Podgornik}},
  \bibinfo{journal}{J. Chem. Phys.} \textbf{\bibinfo{volume}{91}},
  \bibinfo{pages}{5840} (\bibinfo{year}{1989}).

\bibitem[{\citenamefont{Netz and Orland}(2000)}]{NO:EPJE:2000}
\bibinfo{author}{\bibfnamefont{R.~R.} \bibnamefont{Netz}} \bibnamefont{and}
  \bibinfo{author}{\bibfnamefont{H.}~\bibnamefont{Orland}},
  \bibinfo{journal}{Eur. Phys. J. E} \textbf{\bibinfo{volume}{1}},
  \bibinfo{pages}{203} (\bibinfo{year}{2000}).

\bibitem[{\citenamefont{Netz and Orland}(2003)}]{NO:EPJE:2003}
\bibinfo{author}{\bibfnamefont{R.~R.} \bibnamefont{Netz}} \bibnamefont{and}
  \bibinfo{author}{\bibfnamefont{H.}~\bibnamefont{Orland}},
  \bibinfo{journal}{Eur. Phys. J. E} \textbf{\bibinfo{volume}{11}},
  \bibinfo{pages}{301} (\bibinfo{year}{2003}).

\bibitem[{\citenamefont{Bonthuis et~al.}(2011)\citenamefont{Bonthuis, Gekle,
  and Netz}}]{BGN:PRL:2011}
\bibinfo{author}{\bibfnamefont{D.~J.} \bibnamefont{Bonthuis}},
  \bibinfo{author}{\bibfnamefont{S.}~\bibnamefont{Gekle}}, \bibnamefont{and}
  \bibinfo{author}{\bibfnamefont{R.~R.} \bibnamefont{Netz}},
  \bibinfo{journal}{Phys. Rev. Lett.} \textbf{\bibinfo{volume}{107}},
  \bibinfo{pages}{166102} (\bibinfo{year}{2011}).

\bibitem[{\citenamefont{Bonthuis and Netz}(2013)}]{BN:JPC:13}
\bibinfo{author}{\bibfnamefont{D.~J.} \bibnamefont{Bonthuis}} \bibnamefont{and}
  \bibinfo{author}{\bibfnamefont{R.~R.} \bibnamefont{Netz}},
  \bibinfo{journal}{The J. Phys. Chem. B} \textbf{\bibinfo{volume}{117}},
  \bibinfo{pages}{11397} (\bibinfo{year}{2013}).

\bibitem[{\citenamefont{Wang}(2010)}]{Wang:PRE:2010}
\bibinfo{author}{\bibfnamefont{Z.~G.} \bibnamefont{Wang}},
  \bibinfo{journal}{Phys. Rev. E} \textbf{\bibinfo{volume}{81}},
  \bibinfo{pages}{021501} (\bibinfo{year}{2010}).

\bibitem[{\citenamefont{Cherepanov et~al.}(2003)\citenamefont{Cherepanov,
  Feniouk, Junge, and Mulkidjanian}}]{cherepanov2003bj}
\bibinfo{author}{\bibfnamefont{D.~A.} \bibnamefont{Cherepanov}},
  \bibinfo{author}{\bibfnamefont{B.~A.} \bibnamefont{Feniouk}},
  \bibinfo{author}{\bibfnamefont{W.}~\bibnamefont{Junge}}, \bibnamefont{and}
  \bibinfo{author}{\bibfnamefont{A.~Y.} \bibnamefont{Mulkidjanian}},
  \bibinfo{journal}{Biophys. J.} \textbf{\bibinfo{volume}{85}},
  \bibinfo{pages}{1307} (\bibinfo{year}{2003}).

\bibitem[{\citenamefont{Avdeev and Martynov}(1986)}]{AM:CJU:1986}
\bibinfo{author}{\bibfnamefont{S.~M.} \bibnamefont{Avdeev}} \bibnamefont{and}
  \bibinfo{author}{\bibfnamefont{G.~A.} \bibnamefont{Martynov}},
  \bibinfo{journal}{Colloid J. USSR} \textbf{\bibinfo{volume}{48}},
  \bibinfo{pages}{535} (\bibinfo{year}{1986}).

\bibitem[{\citenamefont{Buyukdagli et~al.}(2012)\citenamefont{Buyukdagli,
  Achim, and Ala-Nissila}}]{BAA:JCP:2012}
\bibinfo{author}{\bibfnamefont{S.}~\bibnamefont{Buyukdagli}},
  \bibinfo{author}{\bibfnamefont{C.~V.} \bibnamefont{Achim}}, \bibnamefont{and}
  \bibinfo{author}{\bibfnamefont{T.}~\bibnamefont{Ala-Nissila}},
  \bibinfo{journal}{J. Chem. Phys.} \textbf{\bibinfo{volume}{137}},
  \bibinfo{pages}{104902} (\bibinfo{year}{2012}).

\bibitem[{\citenamefont{Yaroshchuk}(2000)}]{Yaroshchuk:AIS:2000}
\bibinfo{author}{\bibfnamefont{A.~E.} \bibnamefont{Yaroshchuk}},
  \bibinfo{journal}{Adv.Colloid Interface Sci.} \textbf{\bibinfo{volume}{85}},
  \bibinfo{pages}{193} (\bibinfo{year}{2000}).

\bibitem[{\citenamefont{Lau et~al.}(2002)\citenamefont{Lau, Lukatsky, Pincus,
  and Safran}}]{LLP+:PRE:2002}
\bibinfo{author}{\bibfnamefont{A.~W.~C.} \bibnamefont{Lau}},
  \bibinfo{author}{\bibfnamefont{D.~B.} \bibnamefont{Lukatsky}},
  \bibinfo{author}{\bibfnamefont{P.}~\bibnamefont{Pincus}}, \bibnamefont{and}
  \bibinfo{author}{\bibfnamefont{S.~A.} \bibnamefont{Safran}},
  \bibinfo{journal}{Phys. Rev. E} \textbf{\bibinfo{volume}{65}},
  \bibinfo{pages}{051502} (\bibinfo{year}{2002}).

\bibitem[{\citenamefont{Buff and Stillinger}(1963)}]{Buff:JCP:63}
\bibinfo{author}{\bibfnamefont{F.~P.} \bibnamefont{Buff}} \bibnamefont{and}
  \bibinfo{author}{\bibfnamefont{F.~H.} \bibnamefont{Stillinger}},
  \bibinfo{journal}{J. Chem. Phys.} \textbf{\bibinfo{volume}{39}},
  \bibinfo{pages}{1911} (\bibinfo{year}{1963}).

\bibitem[{\citenamefont{Xu and Maggs}(2013)}]{XuMaggs:arXiv:13}
\bibinfo{author}{\bibfnamefont{Z.}~\bibnamefont{Xu}} \bibnamefont{and}
  \bibinfo{author}{\bibfnamefont{A.~C.} \bibnamefont{Maggs}},
  \bibinfo{journal}{J. Comput. Phys.}  (\bibinfo{year}{2014}), \eprint{http://dx.doi.org/10.1016/j.jcp.2014.07.004}

\bibitem[{\citenamefont{Lin et~al.}(2011{\natexlab{a}})\citenamefont{Lin, Yang,
  Meza, Lu, Ying, and E}}]{LYM+:ATMS:2011}
\bibinfo{author}{\bibfnamefont{L.}~\bibnamefont{Lin}},
  \bibinfo{author}{\bibfnamefont{C.}~\bibnamefont{Yang}},
  \bibinfo{author}{\bibfnamefont{J.~C.} \bibnamefont{Meza}},
  \bibinfo{author}{\bibfnamefont{J.}~\bibnamefont{Lu}},
  \bibinfo{author}{\bibfnamefont{L.}~\bibnamefont{Ying}}, \bibnamefont{and}
  \bibinfo{author}{\bibfnamefont{W.}~\bibnamefont{E}}, \bibinfo{journal}{ACM
  Trans. Math. Softw.} \textbf{\bibinfo{volume}{37}}, \bibinfo{pages}{40:1}
  (\bibinfo{year}{2011}{\natexlab{a}}).

\bibitem[{\citenamefont{Lin et~al.}(2011{\natexlab{b}})\citenamefont{Lin, Yang,
  Lu, Ying, and E}}]{LYL+:SJoSC:2011}
\bibinfo{author}{\bibfnamefont{L.}~\bibnamefont{Lin}},
  \bibinfo{author}{\bibfnamefont{C.}~\bibnamefont{Yang}},
  \bibinfo{author}{\bibfnamefont{J.}~\bibnamefont{Lu}},
  \bibinfo{author}{\bibfnamefont{L.}~\bibnamefont{Ying}}, \bibnamefont{and}
  \bibinfo{author}{\bibfnamefont{W.}~\bibnamefont{E}}, \bibinfo{journal}{SIAM
  J. Sci. Comput.} \textbf{\bibinfo{volume}{33}}, \bibinfo{pages}{1329}
  (\bibinfo{year}{2011}{\natexlab{b}}),
  \eprint{http://epubs.siam.org/doi/pdf/10.1137/09077432X}.

\bibitem[{\citenamefont{Pasquali and Maggs}(2009)}]{PM:PRA:2009}
\bibinfo{author}{\bibfnamefont{S.}~\bibnamefont{Pasquali}} \bibnamefont{and}
  \bibinfo{author}{\bibfnamefont{A.~C.} \bibnamefont{Maggs}},
  \bibinfo{journal}{Phys. Rev. A} \textbf{\bibinfo{volume}{79}},
  \bibinfo{pages}{020102} (\bibinfo{year}{2009}).

\bibitem[{\citenamefont{George}(1973)}]{George:SJNA:1973}
\bibinfo{author}{\bibfnamefont{A.}~\bibnamefont{George}},
  \bibinfo{journal}{SIAM J. Numer. Anal.} \textbf{\bibinfo{volume}{10}},
  \bibinfo{pages}{345} (\bibinfo{year}{1973}).

\bibitem[{\citenamefont{Davis}(2006)}]{Davis:SIAM:06}
\bibinfo{author}{\bibfnamefont{T.~A.} \bibnamefont{Davis}},
  \emph{\bibinfo{title}{Direct Methods for Sparse Linear Systems}}
  (\bibinfo{publisher}{SIAM}, \bibinfo{address}{Philadelphia},
  \bibinfo{year}{2006}).

\bibitem[{\citenamefont{Gan and Xu}(2011)}]{GX:PRE:2011}
\bibinfo{author}{\bibfnamefont{Z.}~\bibnamefont{Gan}} \bibnamefont{and}
  \bibinfo{author}{\bibfnamefont{Z.}~\bibnamefont{Xu}}, \bibinfo{journal}{Phys.
  Rev. E} \textbf{\bibinfo{volume}{84}}, \bibinfo{pages}{016705}
  (\bibinfo{year}{2011}).

\bibitem[{\citenamefont{Wang and Ma}(2009)}]{WM:JCP:2009}
\bibinfo{author}{\bibfnamefont{Z.~Y.} \bibnamefont{Wang}} \bibnamefont{and}
  \bibinfo{author}{\bibfnamefont{Y.~Q.} \bibnamefont{Ma}}, \bibinfo{journal}{J.
  Chem. Phys.} \textbf{\bibinfo{volume}{131}}, \bibinfo{pages}{244715}
  (\bibinfo{year}{2009}).

\end{thebibliography}

\end{document}